\documentclass[prd,showpacs,superscriptaddress,nofootinbib,floatfix,11pt]{revtex4-2}
\usepackage[utf8]{inputenc}
\usepackage{amsmath,amssymb}
\usepackage{slashed}
\usepackage{subfigure}
\usepackage[colorlinks=true,
breaklinks=true,
urlcolor=magenta,
citecolor=blue]{hyperref}
\usepackage[usenames,dvipsnames]{color}
\usepackage{appendix}
\usepackage{braket,bm}
\usepackage{multirow}
\usepackage{enumitem}
\usepackage{color}
\usepackage{slashed}
\usepackage{orcidlink}
\usepackage{graphicx}
\usepackage{tikz}
\usepackage{booktabs}
\usepackage{array}
\usepackage{overpic}
\usepackage{float}
\usepackage{threeparttable}
\allowdisplaybreaks[4]

\newcommand{\itp}{\affiliation{CAS Key Laboratory of Theoretical Physics, Institute of Theoretical Physics,\\ Chinese Academy of Sciences, Beijing 100190, China}}

\newcommand{\ucas}{\affiliation{School of Physical Sciences, University of Chinese Academy of Sciences, Beijing 100049, China}}

\newcommand{\peng}{\affiliation{Peng Huanwu Collaborative Center for Research and Education, Beihang University, Beijing 100191, China}}

\newcommand{\scnt}{\affiliation{Southern Center for Nuclear-Science Theory, Institute of Modern Physics, Huizhou 516000, China}}

\newcommand{\hiskp}{\affiliation{Helmholtz-Institut f\"{u}r Strahlen- und Kernphysik and Bethe Center for
Theoretical Physics, \\Universit\"{a}t~Bonn,  D-53115~Bonn,~Germany}}

\newcommand{\BB}{{\mathbb B}}

\begin{document}
\title{$\sigma$ exchange in the one-boson exchange model involving the ground state octet baryons} 
\author{Bing Wu\orcidlink{0009-0004-8178-3015}}\email{wubing@itp.ac.cn}
\itp\ucas
\author{Xiong-Hui Cao\orcidlink{0000-0003-1365-7178}}\email{xhcao@itp.ac.cn}
\itp
\author{Xiang-Kun Dong\orcidlink{0000-0001-6392-7143}}\email{xiangkun@hiskp.uni-bonn.de}
\hiskp
\author{Feng-Kun Guo\orcidlink{0000-0002-2919-2064}}\email{fkguo@itp.ac.cn}
\itp\ucas\peng \scnt

\begin{abstract}
Based on the one-boson-exchange framework that the $\sigma$ meson serves as an effective parameterization for the correlated scalar-isoscalar $\pi\pi$ interaction, we calculate the coupling constants of the $\sigma$ to the  $\frac{1}{2}^+$ ground state light baryon octet $\BB$ by matching the amplitude of $\BB\bar{\BB}\to\pi\pi\to\bar{\BB}\BB$ to that of $\BB\bar{\BB}\to\sigma\to\bar{\BB}\BB$. The former is calculated using a dispersion relation, supplemented with chiral perturbation theory results for the $\BB\BB\pi\pi$ couplings and the Muskhelishvili-Omn\` es representation for the $\pi\pi$ rescattering. 
Explicitly, the coupling constants are obtained as $g_{NN\sigma}=8.7_{-1.9}^{+1.7}$, $g_{\Sigma\Sigma\sigma}=3.5_{-1.3}^{+1.8}$, $g_{\Xi\Xi\sigma}=2.5_{-1.4}^{+1.5}$, and $g_{\Lambda\Lambda\sigma}=6.8_{-1.7}^{+1.5}$. 
These coupling constants can be used in the one-boson-exchange model calculations of the interaction of light baryons with other hadrons.

\end{abstract}

\maketitle
\newpage

\section{Introduction}

In the past few decades, the observation of exotic hadronic states, which cannot be accounted for by the conventional quark model, has propelled the study of exotic states to the forefront of hadron physics; see Refs.~\cite{Hosaka:2016pey,Richard:2016eis,Lebed:2016hpi,Esposito:2016noz,Ali:2017jda,Olsen:2017bmm,Guo:2017jvc,Brambilla:2019esw,Yamaguchi:2019vea,Dong:2021juy,Dong:2021bvy,Chen:2022asf,Meng:2022ozq,Mai:2022eur} for recent reviews on the experimental and theoretical status.  
Hadronic molecules~\cite{Guo:2017jvc}, one of the most promising candidates for exotic states, are loosely bound states of hadrons and a natural extension of the atomic nuclei (such as the deuteron as a proton-neutron bound state) and offer an explanation of the many experimentally observed near-threshold structures, in particular in the heavy-flavor hadron mass region~\cite{Dong:2020hxe}.

As a generalization of the one-pion-exchange potential~\cite{Yukawa:1935xg}, the one-boson-exchange (OBE) model has played a crucial role in studying composite systems of hadrons~\cite{Durso:1977ek,Machleidt:1987hj,Tornqvist:1993ng,Ding:2009vj,CalleCordon:2009pit,Sun:2011uh,Zhao:2013ffn,Liu:2018bkx,Liu:2019stu,Dong:2021juy,Dong:2021bvy}. Taking the deuteron as an example, it is widely accepted in the OBE model that its formation involves the long-range interaction from the one-pion exchange and the middle-range interaction from the $\sigma$-meson exchange; see Ref.~\cite{Machleidt:1987hj} for a detailed review. 
However, unlike narrow width particles that are associated with clear resonance peaks or dips observed in experiments, the scalar-isoscalar $\sigma$ meson, which plays a crucial role in  nuclear and hadron physics, had remained a subject of considerable debates for several decades until it was established as the lowest-lying hadronic resonance in quantum chromodynamics (QCD) in the past twenty years based on rigorous dispersive analyses of $\pi\pi$ scattering~\cite{Zhou:2004ms,Caprini:2005zr,Garcia-Martin:2011nna} (see, e.g., Refs.~\cite{Pelaez:2015qba,Yao:2020bxx} for reviews). The dispersive techniques have recently been applied to determine the nature of the $\sigma$ at unphysical pion masses~\cite{Cao:2023ntr, Rodas:2023twk}.

The $\sigma$ meson in the OBE model can be considered as approximating the correlated $S$-wave isoscalar $\pi\pi$ exchange in a few hundred MeV range~\cite{Durso:1980vn,Machleidt:1987hj,Holzenkamp:1989tq,Meissner:1990kz,Ronchen:2012eg,Reuber:1995vc,Haidenbauer:2005zh}, and some modifications to its properties have been made in order to improve the accuracy of the approximation~\cite{Durso:1980vn,Machleidt:1987hj}. 
However, the effective coupling constants between the $\sigma$ and various hadrons remain highly uncertain. One example is the widely used nucleon-nucleon-$\sigma$ coupling $g_{NN\sigma}$, which ranges roughly from 8 to 14~\cite{Durso:1980vn,Machleidt:1987hj,Ronchen:2012eg}. For its couplings to other ground state octet baryons, $g_{\Sigma\Sigma\sigma}$, $g_{\Xi\Xi\sigma}$, and $g_{\Lambda\Lambda\sigma}$, there are rare systematic discussions and error analyses. Most of them are estimated either by the quenched quark model or the SU(3) symmetry model assuming the $\sigma$ to be a certain member of the light-flavor multiplet~\cite{Liu:2018bkx,Zhao:2013ffn}.
The use of the one-$\sigma$ exchange instead of the correlated $\pi\pi$ exchange may raise some questions: Is this approximation reasonable? How good is the approximation? 
In the present work, we try to address these questions by considering the scattering of the baryon and antibaryon in the ground state octet, $\BB\bar{\BB}\to\bar{\BB}\BB$, through the intermediate state of the correlated $IJ=00$ $\pi\pi$ pair or the $\sigma$ meson and calculate the effective coupling constants of $g_{\BB\BB\sigma}$. 
We will make use of dispersion relations following Ref.~\cite{Donoghue:2006rg}, and similar methods have been used in, e.g., Refs.~\cite{Reuber:1995vc,Haidenbauer:2005zh} to derive the baryon-baryon-$\sigma$ couplings, and Ref.~\cite{Kim:2019rud} to derive the $\sigma$ coupling to heavy mesons.
Here, we will match the dispersive amplitudes of $\BB \bar\BB \to \pi\pi$ at low energies to the chiral amplitudes up to the next-to-leading order (NLO).

This paper is structured as follows. The formalism is presented in Sec.~\ref{formalism}, including the calculation of the OBE amplitude in Sec.~\ref{OBE amplitude} and the amplitude from the dispersion relation (DR) with a careful treatment of kinematical singularities in Sec.~\ref{DR amplitude}. In Sec.~\ref{numerical result}, we conduct an analysis of the two amplitudes and present the scalar coupling constants $g_{\BB\BB\sigma}$ along with an error analysis. This includes a comparison of the $s$-channel processes, as detailed in Sec.~\ref{s-channel simulation}, and a comparison of the corresponding $t/u$-channel processes utilizing the crossing symmetry in Sec.~\ref{t/u-channel simulation}. 
A brief summary is given in Sec.~\ref{summary}. The adopted conventions, the result of $NN\sigma$ coupling in the SU(2) framework, and crossing symmetry relations are relegated to the appendices.

\section{Formalism}\label{formalism}

\begin{figure}[tb]
\centering
    \begin{overpic}[width=0.8\linewidth]{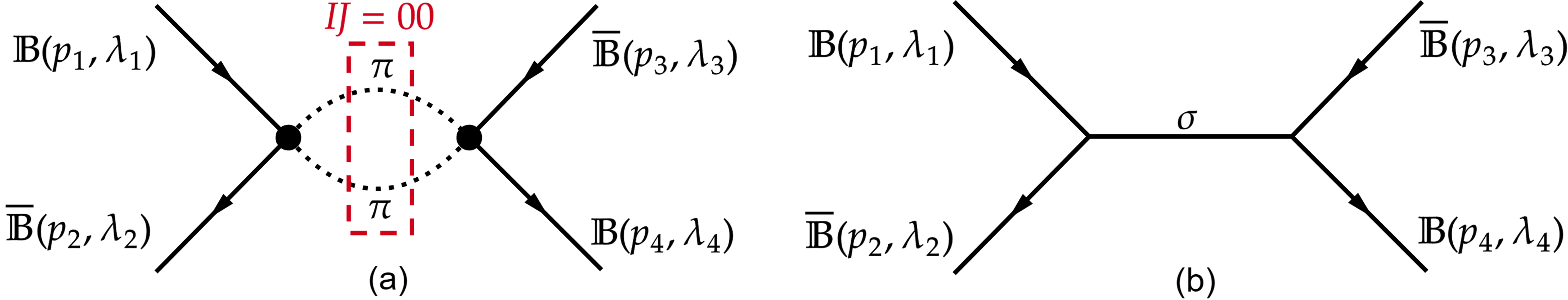}
    \end{overpic}
\caption{Feynman diagrams for the $s$-channel process of $\BB\bar{\BB}\to \bar{\BB}\BB$ with the intermediate state of $\pi\pi$ (a) or $\sigma$ (b). In (a), the black dots imply that the $\pi\pi$ rescattering is included.}
  \label{FIG1}
\end{figure}

To determine the scalar coupling, $g_{\BB\BB\sigma}$, between the baryon $\BB$ in the $\frac{1}{2}^+$ ground baryon octet and the $\sigma$ meson, we first utilize the DR and chiral perturbation theory (ChPT) to calculate the amplitude of $\BB(p_1,\lambda_1)+\bar{\BB}(p_2,\lambda_2)\to \bar{\BB}(p_3,\lambda_3)+\BB(p_4,\lambda_4)$ with a correlated $\pi\pi$ intermediate state, as depicted in Fig.~\ref{FIG1}(a). Here, $p_i$ and $\lambda_i$ represent the four momentum and the helicity of particle $\BB$ or $\bar{\BB}$, respectively. Additionally, we restrict the quantum numbers of the two-body intermediate state, $\pi\pi$, to be $IJ=00$. This $S$-wave amplitude can be denoted as $\mathcal{M}^{\rm{DR}}_{\BB(\lambda_1)+\bar{\BB}(\lambda_2)\to\bar{\BB}(\lambda_3)+\BB(\lambda_4),0}(s)$, where the subscript 0 indicates the $S$-wave, $s$ represents the square of the total energy of the system in the center-of-mass (c.m.) frame,\footnote{For an $s$-channel process of $\BB(p_1)+\bar{\BB}(p_2)\to\bar{\BB}(p_3)+\BB(p_4)$, as illustrated in Fig.~\ref{FIG1}, $s=(p_1+p_2)^2$ while $t=(p_1-p_3)^2$.} and the superscript DR indicates the result obtained from the DR. Next, we proceed to calculate the same amplitude, with the intermediate $\sigma$ meson, as depicted in Fig.~\ref{FIG1}(b). In this case, we utilize the OBE model, and the corresponding amplitude can be expressed as $\mathcal{M}^{\rm{OBE}}_{\BB(\lambda_1)+\bar{\BB}(\lambda_2)\to\bar{\BB}(\lambda_3)+\BB(\lambda_4)}(s)$. Finally, we compare the aforementioned amplitudes
to extract the coupling constant $g_{\BB\BB\sigma}$ in the phenomenological baryon-baryon-$\sigma$ coupling Lagrangian.

\subsection{The OBE amplitude}\label{OBE amplitude}

According to the following effective Lagrangian coupling the $\sigma$ meson to the baryons in the SU(3) flavor octet~\cite{Machleidt:1987hj,Yalikun:2021dpk},
\begin{align}
\mathcal{L}_{\Sigma\Sigma\sigma}&=-g_{\Sigma\Sigma\sigma}\left( \bar{\Sigma}^+\Sigma^- + \bar{\Sigma}^0\Sigma^0 + \bar{\Sigma}^-\Sigma^+ \right)\sigma\,,\nonumber\\
\mathcal{L}_{\Xi\Xi\sigma}&=-g_{\Xi\Xi\sigma}\left( \bar{\Xi}^0\Xi^0 +\bar{\Xi}^+\Xi^- \right)\sigma\,,\nonumber\\
\mathcal{L}_{\Lambda\Lambda\sigma}&=-g_{\Lambda\Lambda\sigma}\bar{\Lambda}\Lambda\sigma\,,\nonumber\\
\mathcal{L}_{NN\sigma}&=-g_{NN\sigma}(\bar{p}p+\bar{n}n)\sigma\,,
\end{align}
the OBE amplitude for the Feynman diagram depicted in Fig.~\ref{FIG1}(b)
reads
\begin{align}
\mathcal{M}^{\rm{OBE}}_{\BB(\lambda_1)+\bar{\BB}(\lambda_2)\to\bar{\BB}(\lambda_3)+\BB(\lambda_4)}=C_\BB g_{\BB\BB\sigma}^2\frac{\bar{v}^{\lambda_2}(p_2) u^{\lambda_1}(p_1) \bar{u}^{\lambda_4}(p_4) v^{\lambda_3}(p_3)}{(p_1+p_2)^2-m_\sigma^2}\,,
\end{align}
where $C_\BB$ is a flavor factor, $C_\Sigma=3$, $C_\Xi=-2$, $C_\Lambda=1$ and $C_N=-2$. For simplicity, we choose $\lambda_i={1}/{2}$ $(i=1,\ldots,4)$ throughout the paper.\footnote{Other choices, e.g., physical amplitudes using the orbital-spin basis are also accessible. The final results do not depend on the choice.}
With this choice we have
\begin{align}
\mathcal{M}^{\rm{OBE}}_{\BB\bar{\BB}\to\bar{\BB}\BB}(s)=C_\BB g_{\BB \BB\sigma}^2\frac{s-4m_\BB^2}{s-m_\sigma^2}\,,\label{equ:OBE amplitude}
\end{align}
where $m_\BB$ is the isospin averaged mass of the baryon $\BB$.\footnote{Since we are not interested in the isospin symmetry breaking effects, the isospin averaged mass is used for all particles within the same isospin multiplet.} 

\subsection{The dispersive representation}\label{DR amplitude}

\subsubsection{The DR and the kinematical singularity}\label{DR UR KS}
 
One can write down a dispersive representation of the $\BB \bar \BB$ scattering amplitude corresponding to Fig.~\ref{FIG1}(a) as
\begin{align}
\mathcal{M}^{\rm{DR}}_{\BB \bar{\BB}\to\bar{\BB}\BB,0}(s)=\frac{s-4m_\BB^2}{2\pi i}\int_{4M_\pi^2}^{+\infty}\frac{{\rm{disc}}\left[
\mathcal{M}^{\rm{DR}}_{\BB \bar{\BB}\to\bar{\BB}\BB,0}(z)
\right]}{(z-s)(z-4m_\BB^2)} {\rm{d}}z\,.\label{DR integral}
\end{align}
Here a once-subtracted dispersive integral is employed to facilitate the convergence of the dispersive integral. The threshold $s=4m_\BB^2$ is chosen as the subtraction point, and we set the subtraction constant $\mathcal{M}^{\rm{DR}}(s=4m_\BB^2)=0$ as Eq.~(\ref{equ:OBE amplitude}) since the two amplitudes will be matched later.  

In order to capture the $\pi\pi$ rescattering in the $\sigma$ region and avoid the interference from other resonances, e.g., the $f_0(980)$, the upper limit of the dispersive integral in Eq.~\eqref{DR integral} is set to $s_0 = (0.8\ {\rm{GeV}})^2$, as in Ref.~\cite{Donoghue:2006rg} (see also Ref.~\cite{Gasser:1990bv}). 
We will investigate the impact of varying the upper limit of the integration on the final result and regard it as a part of the uncertainty of the coupling constants.

Next, let us discuss the discontinuity. Taking into account the unitary relation that is fulfilled by the partial-wave $T$-matrix elements, we can express the discontinuity of the $S$-wave amplitude (here the partial wave refers to that between the pions) along the cut $s\in [4M_\pi^2,+\infty)$ in terms of $T_{\BB\bar{\BB}\to\pi\pi,0}(s)$ and $T_{\pi\pi\to \bar{\BB}\BB,0}(s)$. However, it is crucial to notice that when dealing with systems that involve spins, particularly those containing fermions, kinematical singularities arise~\cite{Martin:1970hmp}. These singularities stem from the definition of the wave functions for the initial and final states.
Following Ref.~\cite{Martin:1970hmp}, we introduce the kinematical-singularity-free amplitudes,
\begin{align}
    T^{\rm{new}}_{\BB\bar{\BB}\to\pi\pi,0}(s)=\sqrt{s-4m_\BB^2}\, T_{\BB\bar{\BB}\to\pi\pi,0}(s)\label{equ:new amplitude1}\,,\\
    T^{\rm{new}}_{\pi\pi\to\bar{\BB}\BB,0}(s)=\sqrt{s-4m_\BB^2}\, T_{\pi\pi\to\bar{\BB}\BB,0}(s)\,.
    \label{equ:new amplitude2}
\end{align}
Then the unitary relation for the $S$-wave $T$-matrix elements is given by
\begin{align}
 {\rm{disc}}\left[ \mathcal{M}_{\BB\bar{\BB}\to\bar{\BB}\BB,0}(s) \right]=2i\rho_\pi(s)\frac{T_{\BB\bar{\BB}\to\pi\pi,0}^{\rm{new}}(s)T^{{\rm{new}}\ *}_{\pi\pi\to \bar{\BB}\BB,0}(s)}{s-4m_\BB^2}\theta\left( \sqrt{s}-2M_\pi \right) ,
\end{align}
where $\rho_\pi(s)=\frac{1}{16\pi}\sqrt{\frac{s-4M_\pi^2}{s}}$ is the two-body phase space factor. Moreover, as we will discuss in detail in Sec.~\ref{The FSI}, the treatment of kinematical singularity plays a vital role in guaranteeing the self-consistency of the theory.

Furthermore, with the phase conventions outlined in Appendix~\ref{SecI} and considering the isospin scalar $\pi\pi$ system, we obtain the following relations
\begin{gather}
    T_{\Sigma\bar{\Sigma}\to\pi\pi,0}(s)=-T_{\pi\pi\to\bar{\Sigma}\Sigma,0}(s)\,,\nonumber\\
    T_{\Xi\bar{\Xi}\to\pi\pi,0}(s)=T_{\pi\pi\to\bar{\Xi}\Xi,0}(s)\,,\nonumber\\
    T_{\Lambda\bar{\Lambda}\to\pi\pi,0}(s)=-T_{\pi\pi\to\bar{\Lambda}\Lambda,0}(s)\,,\nonumber\\
    T_{N \bar{N}\to\pi\pi,0}(s)=T_{\pi\pi\to\bar{N}N,0}(s)\,.
\end{gather}
Then, we obtain the following discontinuities,
\begin{gather}
    {\rm{disc}}\left[ \mathcal{M}_{\Sigma\bar{\Sigma}\to\bar{\Sigma}\Sigma,0}(s) \right]=2i\rho_{\pi}\frac{-\left| T_{\Sigma\bar{\Sigma}\to\pi\pi,0}^{\rm{new}}(s) \right|^2}{s-4m_\Sigma^2}\theta\left( \sqrt{s}-2M_\pi \right) ,\label{equ:disc of Sigma}\\
    {\rm{disc}}\left[ \mathcal{M}_{\Xi\bar{\Xi}\to\bar{\Xi}\Xi,0}(s) \right]=2i\rho_{\pi}\frac{\left| T_{\Xi\bar{\Xi}\to\pi\pi,0}^{\rm{new}}(s) \right|^2}{s-4m_\Xi^2}\theta\left( \sqrt{s}-2M_\pi \right),\label{equ:disc of Xi}\\
    {\rm{disc}}\left[ \mathcal{M}_{\Lambda\bar{\Lambda}\to\bar{\Lambda}\Lambda,0}(s) \right]=2i\rho_{\pi}\frac{-\left| T_{\Lambda\bar{\Lambda}\to\pi\pi,0}^{\rm{new}}(s) \right|^2}{s-4m_\Lambda^2}\theta\left( \sqrt{s}-2M_\pi \right) ,\label{equ:disc of Lambda}\\
    {\rm{disc}}\left[ \mathcal{M}_{N\bar{N}\to\bar{N}N,0}(s) \right]=2i\rho_{\pi}\frac{\left| T_{N\bar{N}\to\pi\pi,0}^{\rm{new}}(s) \right|^2}{s-4m_N^2}\theta\left( \sqrt{s}-2M_\pi \right).\label{equ:disc of N}    
\end{gather}

\subsubsection{The SU(3) ChPT framework}

\begin{figure}[tbh]
\centering
    \begin{overpic}[width=0.7\linewidth]{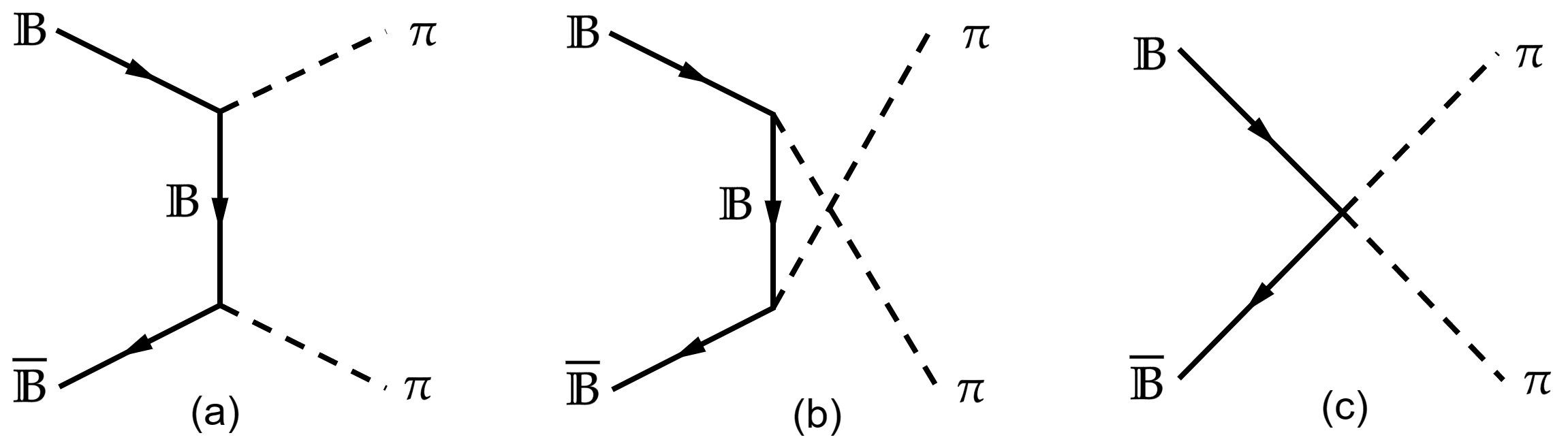}
    \end{overpic}
\caption{The tree-level Feynman diagrams for the process of $\BB\bar{\BB}\to\pi\pi$. }
  \label{FIG2} 
\end{figure}

To obtain the low-energy $S$-wave amplitudes $T_{\BB\bar{\BB}\to\pi\pi,0}(s)$, we need the corresponding chiral baryon-meson Lagrangian. The leading-order (LO) chiral Lagrangian is given by~\cite{Krause:1990xc},
\begin{gather}
\mathcal{L}_{\mathbb{M} \BB}^{(1)}=\left\langle \bar{\BB}(i \slashed{\mathcal{D}} -m_0)\BB \right\rangle+\frac{D}{2}\left\langle \bar{\BB}\gamma^\mu\gamma_5 \{ u_\mu,\BB \} \right\rangle+\frac{F}{2}\left\langle\bar{\BB}\gamma^\mu\gamma_5 \left[ u_\mu,\BB \right] \right\rangle ,  \label{LO Lagrangian}
\end{gather}
which contains three low-energy constants (LECs), $m_0$, $D$ and $F$. 
Here, $\langle \cdot \rangle$ means trace in the flavor space, the baryon octet are collected in the matrix,
\begin{align}
  \BB = \left(\begin{array}{ccc}
    \frac{1}{\sqrt{2}} \Sigma^0+\frac{1}{\sqrt{6}} \Lambda & \Sigma^{+} & p \\
    \Sigma^{-} & -\frac{1}{\sqrt{2}} \Sigma^0+\frac{1}{\sqrt{6}} \Lambda & n \\
    \Xi^{-} & \Xi^0 & -\frac{2}{\sqrt{6}} \Lambda
    \end{array}\right),
\end{align}
and the chiral vielbein and covariant derivative are given by 
\begin{align}
  u_\mu = i u^{\dagger} \partial_\mu u+i u \partial_\mu u^{\dagger}, \quad \mathcal{D}_\mu \BB = \partial_\mu B+\left[\Gamma_\mu, B\right], \quad \Gamma_\mu = \frac{1}{2} \left(u^{\dagger}\partial_\mu u + u\partial_\mu u^\dagger\right),
\end{align}
with $u^2= U$, $U = \exp \left(i {\sqrt{2}\Phi}/{F_\pi}\right)$, where 
\begin{align}
  \Phi = \left(\begin{array}{ccc}
    \frac{\pi^0}{\sqrt{2}}+\frac{\eta}{\sqrt{6}} & \pi^{+} & K^{+} \\
    \pi^{-} & -\frac{\pi^0}{\sqrt{2}}+\frac{\eta}{\sqrt{6}} & K^0 \\
    K^{-} & \bar{K}^0 & -\frac{2}{\sqrt{6}} \eta
    \end{array}\right) .
\end{align}
Notice that the chiral connection $\Gamma_\mu$ containing two pions is a vector, the two pions from that term cannot be in the $S$-wave. 
As a result, only the $t$- and $u$-channel exchanges depicted in Fig.~\ref{FIG2} (a) and (b), which contribute to the LHC part of $T_{\BB\bar{\BB}\to\pi\pi,0}(s)$, will be present in the LO calculation. In addition, the LO Lagrangian contains the $\Sigma\Lambda\pi$ coupling terms of the form $\bar{\Lambda}\gamma^\mu\gamma_5\partial_\mu\pi\Sigma$ and $\bar{\Sigma}\gamma^\mu\gamma_5\partial_\mu\pi\Lambda$. 
Therefore, it is necessary to consider the exchange of $\Sigma$ in the $\Lambda\bar{\Lambda}\to\pi\pi$ process and the exchange of $\Lambda$ in the $\Sigma\bar{\Sigma}\to\pi\pi$ process.

The $\mathcal{O}(p^2)$ chiral Lagrangian contains the $\bar{\BB}\BB\pi\pi$ contact contribution with the $S$-wave pion pair, as illustrated in Fig.~\ref{FIG2} (c). 
At the NLO, the number of LECs increases, and the Lagrangian reads~\cite{Frink:2004ic,Oller:2006yh},\footnote{Here we use the Lagrangian in \cite{Oller:2006yh} while the one in Ref.~\cite{Frink:2004ic} has redundant terms. Note also the ordering of the operators in Ref.~\cite{Oller:2006yh} is different from that in Refs.~\cite{Frink:2004ic,Mai:2012dt}.}
\begin{align}
    \mathcal{L}_{\mathbb{MB}}^{(2)}=&\, b_D \langle \bar{\BB} \{ \chi_+,\BB \} \rangle+b_F \langle \bar{\BB} \left[ \chi_+,\BB \right] \rangle+b_0 \langle \bar{\BB}\BB \rangle \langle \chi_+ \rangle\notag\\
    &+b_1 \langle \bar{\BB} \left[ u^\mu ,\left[ u_\mu,\BB\right] \right] \rangle+b_2 \langle  \bar{\BB} \{ u^\mu ,\{ u_\mu ,\BB \} \} \rangle\notag\\
    &+b_3 \langle \bar{\BB} \{ u^\mu, \left[ u_\mu,\BB\right] \} \rangle+b_4\langle \bar{\BB}\BB \rangle\langle u^\mu u_\mu \rangle\notag\\
    &+ib_5 \left( \langle \bar{\BB} \left[ u^\mu,\left[ u^\nu,\gamma_\mu \mathcal{D}_\nu \BB\right]\right] \rangle - \langle \bar{\BB} \overleftarrow{\mathcal{D}}_\nu \left[ u^\nu,\left[ u^\mu,\gamma_\mu \BB\right] \right] \rangle \right) \label{NLO Lagrangian}\\
    &+ib_6 \left( \langle \bar{\BB} \left[ u^\mu,\{ u^\nu,\gamma_\mu \mathcal{D}_\nu \BB\} \right] \rangle - \langle \bar{\BB} \overleftarrow{\mathcal{D}}_\nu \{ u^\nu,\left[ u^\mu,\gamma_\mu \BB\right] \} \rangle \right)\notag\\
    &+ib_7 \left( \langle \bar{\BB} \{ u^\mu,\{ u^\nu,\gamma_\mu \mathcal{D}_\nu \BB\} \} \rangle - \langle \bar{\BB} \overleftarrow{\mathcal{D}}_\nu \{ u^\nu,\{ u^\mu,\gamma_\mu \BB\} \} \rangle \right)\notag\\
    &+ib_8\left( \langle \bar{\BB}\gamma_\mu \mathcal{D}_\nu \BB \rangle-\langle \bar{\BB}\overleftarrow{\mathcal{D}}_\nu\gamma_\mu \BB \rangle \right) \langle u^\mu u^\nu \rangle+\raisebox{.5ex}{\ldots} ,\notag
\end{align}
where $\chi_{ \pm}=u^{\dagger} \chi u^{\dagger} \pm u \chi^{\dagger} u, \chi=2B_0 \mathcal{M}$ with $B_0$ a constant related to the quark condensate in the chiral limit and $\mathcal{M}$ the light-quark mass matrix.
We will use the values of involved LECs from Fit~II in Ref.~\cite{Ren:2012aj}, which are $D=0.8$, $F=0.46$, $b_D=0.222(20)$, $b_F=-0.428(12)$, $b_0=-0.714(21)$, $b_1=0.515(132)$, $b_2=0.148(48)$, $b_3=-0.663(155)$, $b_4=-0.868(105)$, $b_5=-0.643(246)$, $b_6=-0.268(334)$, $b_7=0.176(72)$, $b_8=-0.0694(1638)$.

\subsubsection{The partial-wave amplitudes}

Using the LO and NLO Lagrangians given in Eqs.~(\ref{LO Lagrangian}, \ref{NLO Lagrangian}), we can calculate the tree-level amplitude for the process of $\BB\bar{\BB}\to\pi\pi$ as depicted in Fig.~\ref{FIG2}. However, in order to determine the final state $\pi\pi$ with $IJ=00$, we need to perform a partial-wave (PW) expansion. The generalized PW expansion of the helicity amplitude for arbitrary spin can be found in Ref.~\cite{Jacob:1959at}. The final PW amplitude for $\BB\bar{\BB}\to\pi\pi$ reads 
\begin{align}
    T_{\BB\bar{\BB}\to\pi\pi,L}(s)=\frac{1}{4\pi}\int{\rm{d}}\Omega\sqrt{\frac{4\pi}{2L+1}}Y^*_{L,\lambda_1-\lambda_2}(\theta,\phi)e^{i(\lambda_1-\lambda_2)\phi}\langle \theta,0;0,0\vert \hat{T} \vert 0,0;\lambda_1,\lambda_2\rangle\label{equ:PWA}\,,
\end{align}
where {$L$ is the relative orbital angular momentum of the pions}, $\lambda_1$ and $\lambda_2$ are the third components of the helicities of $\BB$ and $\bar{\BB}$. 
{The basis is such that $\BB\bar \BB$ is expanded in terms of $|\theta_0,\phi_0;\lambda_1,\lambda_2\rangle$, and the $\BB\bar \BB$ relative momentum is chosen to be along the $z$ axis so that $\theta_0=\phi_0=0$; $(\theta,\phi)$ are the polar and azimuthal angles of the $\pi\pi$ relative momentum.}

For the tree-level $S$-wave amplitude for $\BB\bar{\BB}\to\pi\pi$, the LHC part from the $t$- and $u$-channel baryon exchange is\footnote{Here we consider only the baryon exchanges such that the two mesons emitted are two pions since we focus on the correlated $S$-wave two-pion exchange. That is, although we use an SU(3) chiral Lagrangian, the exchanged baryon has the same strangeness as the external ones. The framework may be understood as an SU(2) one for each of the baryons, but with the LECs matched to those in the SU(3) Lagrangian.} 
\begin{align}
  \hat{A}_0^{N}(s)&=-\frac{\sqrt{3}(D+F)^2 m_N}{F_\pi^2 }\frac{L(s,m_N,m_N,M_\pi)}{\sqrt{s-4m_N^2}}\,,\label{equ:LHC amplitude for N}\\
    \hat{A}_0^{\Sigma}(s)&=-\frac{4 \sqrt{2} F^2 m_\Sigma}{F_\pi^2 }\frac{L(s,m_\Sigma,m_\Sigma,M_\pi)}{\sqrt{s-4m_\Sigma^2}}-\frac{\sqrt{2}D^2(m_\Sigma+m_\Lambda)}{3F_\pi^2}\frac{L(s,m_\Sigma,m_\Lambda,M_\pi)}{\sqrt{s-4m_\Sigma^2}}\,,\label{equ:LHC amplitude for Sigma}\\
    \hat{A}_0^{\Lambda}(s)&=\sqrt{\frac{2}{3}}\frac{D^2\left(m_\Lambda+m_\Sigma\right)}{F_\pi^2}\frac{L(s,m_\Lambda,m_\Sigma,M_\pi)}{\sqrt{s-4m_\Lambda^2}}\,,\label{equ:LHC amplitude for Lambda}\\
    \hat{A}_0^{\Xi}(s)&=\frac{\sqrt{3}(D-F)^2 m_\Xi}{F_\pi^2 }\frac{L(s,m_\Xi,m_\Xi,M_\pi)}{\sqrt{s-4m_\Xi^2}}\,,\label{equ:LHC amplitude for Xi}
\end{align}
where
\begin{align} 
L(s,m_1,m_2,m)&=s-2m_1(m_1-m_2)+H_0(s,m_1,m_2,m)H_1(s,m_1,m_2,m) \,,\notag\\
H_0(s,m_1,m_2,m)&=2(m_1+m_2)\left[-2m^2m_1+2m_1(m_1-m_2)^2+m_2s \right] ,\notag\\
H_1(s,m_1,m_2,m)&=\frac{H_2^+(s,m_1,m_2,m)-H_2^-(s,m_1,m_2,m)}{2\sqrt{(s-4m^2)(s-4m_1^2)}}\,,\notag\\
H_2^\pm(s,m_1,m_2,m)&={\rm{ln}}\left[ s-2(m^2+m_1^2-m_2^2) \mp \sqrt{(s-4m^2)(s-4m_1^2)} \right] .\notag
\end{align}

The contact terms, which are from the NLO Lagrangian and contribute to the RHC part of $T_{\BB\bar{\BB}\to\pi\pi,0}(s)$ after taking into account the $\pi\pi$ rescattering, read
\begin{align}
  A_0^{N}(s)=\frac{\sqrt{s-4 m_N^2}}{4 \sqrt{3} F_\pi^2}\Bigl(&8 M_\pi^2(6 b_0-3(b_1+b_2+b_3+2 b_4-b_D-b_F)-2(b_5+b_6+b_7+2 b_8) m_N)\notag\\
    &+4(3(b_1+b_2+b_3+2 b_4)+(b_5+b_6+b_7+2 b_8) m_N) s\Bigr) ,\label{equ:RHC amplitude for N} \\
    A_0^{\Sigma}(s)=\frac{\sqrt{s-4 m_\Sigma^2}}{6 \sqrt{2} F_\pi^2}\Bigl(&8 M_\pi^2(9 b_0-12 b_1-6 b _2-9 b_4+9 b_D-2(4 b_5+2 b_7+3 b_8) m_\Sigma)\notag\\
    &+4(12 b_1+6 b_2+9b_4+4 b_5 m_\Sigma+2 b_7 m_\Sigma+3 b_8 m_\Sigma) s\Bigr) ,\label{equ:RHC amplitude for Sigma}\\
    A_0^{\Lambda}(s)=-\frac{\sqrt{s-4 m_\Lambda^2}}{6 \sqrt{6} F_\pi^2}\Bigl(&8 M_\pi^2(9 b_0-6 b_2-9 b_4+3 b_D-4 b_7 m_\Lambda-6 b_8 m_\Lambda)\notag\\
    &+4(6 b_2+9 b_4+2 b_7 m_\Lambda+3 b_8 m_\Lambda) s\Bigr) ,\label{equ:RHC amplitude for Lambda}\\
    A_0^{\Xi}(s)=-\frac{\sqrt{s-4 m_\Xi^2}}{4 \sqrt{3} F_\pi^2}\Bigl(&8 M_\pi^2(6 b_0-3(b_1+b_2-b_3+2 b_4-b_D+b_F)-2(b_5-b_6+b_7+2 b_8) m_\Xi)\notag\\
    &+4(3(b_1+b_2-b_3+2 b_4)+(b_5-b_6+b_7+2 b_8) m_\Xi) s\Bigr) ,\label{equ:RHC amplitude for Xi}
\end{align}
where the parameter $F_\pi$ is the decay constant of the $\pi$ in the chiral limit. Since we use the LECs determined in Ref.~\cite{Ren:2012aj}, we adopt the same value $F_\pi=87.1$~MeV~\cite{Amoros:2001cp} for consistency. 

Moreover, employing Eq.~(\ref{equ:new amplitude1}), the tree-level $S$-wave amplitudes for $\BB\bar{\BB}\to\pi\pi$ after eliminating the kinematical singularities read, for the LHC part,
\begin{align}
  \hat{A}_0^{N\ {\rm{new}}}(s)&=-\frac{\sqrt{3}(D+F)^2 m_N}{F_\pi^2 }L(s,m_N,m_N,M_\pi)\label{equ:new defined LHC amplitude for N} ,\\
    \hat{A}_0^{\Sigma\ {\rm{new}}}(s)&=-\frac{4 \sqrt{2} F^2 m_\Sigma}{F_\pi^2 }L(s,m_\Sigma,m_\Sigma,M_\pi)-\frac{\sqrt{2}D^2(m_\Sigma+m_\Lambda)}{3F_\pi^2}L(s,m_\Sigma,m_\Lambda,M_\pi)\label{equ:new defined LHC amplitude for Sigma} ,\\
    \hat{A}_0^{\Lambda\ {\rm{new}}}(s)&=\sqrt{\frac{2}{3}}\frac{D^2\left(m_\Lambda+m_\Sigma\right)}{F_\pi^2}L(s,m_\Lambda,m_\Sigma,M_\pi)\label{equ:new defined LHC amplitude for Lambda} ,\\
    \hat{A}_0^{\Xi\ {\rm{new}}}(s)&=\frac{\sqrt{3}(D-F)^2 m_\Xi}{F_\pi^2 }L(s,m_\Xi,m_\Xi,M_\pi)\label{equ:new defined LHC amplitude for Xi} ,
\end{align}
and for the contact term part,
\begin{align}
  A_0^{N\ {\rm{new}}}(s)=\frac{s-4 m_N^2}{4 \sqrt{3} F_\pi^2}\Bigl(&8 M_\pi^2[6 b_0-3(b_1+b_2+b_3+2 b_4-b_D-b_F)-2(b_5+b_6+b_7+2 b_8) m_N]\notag\\
    &+4[3(b_1+b_2+b_3+2 b_4)+(b_5+b_6+b_7+2 b_8) m_N] s\Bigr), \label{equ:new defined RHC amplitude for N}  \\
    A_0^{\Sigma\ {\rm{new}}}(s)=\frac{s-4 m_\Sigma^2}{6 \sqrt{2} F_\pi^2}\Bigl(&8 M_\pi^2 [9 b_0-12 b_1-6 b _2-9 b_4+9 b_D-2(4 b_5+2 b_7+3 b_8) m_\Sigma]\notag\\
    &+4(12 b_1+6 b_2+9b_4+4 b_5 m_\Sigma+2 b_7 m_\Sigma+3 b_8 m_\Sigma) s\Bigr)\label{equ:new defined RHC amplitude for Sigma} ,\\
    A_0^{\Lambda\ {\rm{new}}}(s)=-\frac{s-4 m_\Lambda^2}{6 \sqrt{6} F_\pi^2}\Bigl(&8 M_\pi^2(9 b_0-6 b_2-9 b_4+3 b_D-4 b_7 m_\Lambda-6 b_8 m_\Lambda)\notag\\
    &+4(6 b_2+9 b_4+2 b_7 m_\Lambda+3 b_8 m_\Lambda) s\Bigr)\label{equ:new defined RHC amplitude for Lambda},\\
    A_0^{\Xi\ {\rm{new}}}(s)=-\frac{s-4 m_\Xi^2}{4 \sqrt{3} F_\pi^2}\Bigl(&8 M_\pi^2[6 b_0-3(b_1+b_2-b_3+2 b_4-b_D+b_F)-2(b_5-b_6+b_7+2 b_8) m_\Xi]\notag\\
    &+4[3(b_1+b_2-b_3+2 b_4)+(b_5-b_6+b_7+2 b_8) m_\Xi] s\Bigr)\label{equ:new defined RHC amplitude for Xi} .
\end{align}

\subsubsection{The Muskhelishvili-Omn\`es representation}\label{The FSI}

We now incorporate the $\pi\pi$ rescattering based on the tree-level amplitude, into the  Muskhelishvili-Omn\`es representation. For the $\BB\bar{\BB}\to\pi\pi$ process, we partition the total $S$-wave kinematical-singularity-free amplitude into the LHC and the RHC parts,
\begin{align}   T_{\BB\bar{\BB}\to\pi\pi,0}^{\rm{new}}(s)=R_{\BB,0}^{\rm{new}}(s)+L_{\BB,0}^{\rm{new}}(s) .
\end{align}

Utilizing the $\pi\pi$ amplitude in the  scalar-isoscalar channel $T_{\pi\pi\to\pi\pi,0}(s)=e^{i\delta_0(s)}\sin\delta_0(s)/\rho_\pi(s)$, where $\delta_0(s)$ is the $S$-wave isoscalar phase shift, and since there is no overlap between the LHC and RHC for kinematic-singularity-free amplitudes,\footnote{The RHC is chosen to be along the positive $s$ axis in the interval $s\geq 4M_\pi^2$. The LHC is in the interval $\left(-\infty,\left(4m_\BB^2M_\pi^2-(m_0^2-m_\BB^2-M_\pi^2)^2\right)/{m_0^2}\right]$ for the $t$- or $u$-channel process of $\BB \bar{\BB}\to \pi\pi$, where $m_0$ represents the mass of the exchanged particle. It can be easily proven that $\left(4m_\BB^2M_\pi^2-(m_0^2-m_\BB^2-M_\pi^2)^2\right)/{m_0^2}\leq 4M_\pi^2$. } the unitary relation implies,
\begin{align}
    {\rm{disc}}\left[ R_{\BB,0}^{\rm{new}}(s) \right]=2i\left( R_{\BB,0}^{\rm{new}}(s)+L_{\BB,0}^{\rm{new}}(s) \right)e^{-i\delta_0(s)}\sin\delta_0(s)\theta(\sqrt{s}-2M_\pi) .\label{equ: RHC equation for solve}
\end{align}

To solve this equation, we first define the Omn\`es function~\cite{Omnes:1958hv},
\begin{align}
    \Omega_0(s)\equiv \exp{\left[\frac{s}{\pi}\int_{4M_\pi^2}^{+\infty}\frac{\delta_0(z)}{z(z-s)}{\rm{d}}z\right]} .
\end{align}
By using $\Omega_0(s\pm i\epsilon)=\vert\Omega_0(s)\vert e^{\pm i\delta_0(s)}$, we further derive 
\begin{align}
   {\rm{disc}}\left[ \frac{R_{\BB,0}^{\rm{new}}(s)}{\Omega_0(s)} \right]=2i \frac{L_{\BB,0}^{\rm{new}}(s)}{\vert\Omega_0(s)\vert}\sin\delta_0(s)\theta(\sqrt{s}-2M_\pi) .\label{disc for LHC}
\end{align}
Therefore, we can derive a DR with $n$ subtractions,
\begin{align}
    R_{\BB,0}^{\rm{new}}(s)=\Omega_0(s)\left(P_{n-1}(s)+\frac{s^n}{\pi}\int_{4M_\pi^2}^{+\infty}{\rm{d}}z \frac{L_{\BB,0}^{\rm{new}}(z)\sin\delta_0(z)}{(z-s)z^n\vert \Omega_0(z) \vert}\right) ,\label{equ:RHC general solution}
\end{align}
where $P_{n-1}(s)$ is an arbitrary polynomial of order $n-1$. 
Finally, we obtain a DR for $T_{\BB\bar{\BB}\to\pi\pi,0}^{\rm{new}}(s)$ as 
\begin{align}
T_{\BB\bar{\BB}\to\pi\pi,0}^{\rm{new}}(s)=L_{\BB,0}^{\rm{new}}(s)+\Omega_0(s)\left(P_{n-1}(s)+\frac{s^n}{\pi}\int_{4M_\pi^2}^{+\infty}{\rm{d}}z \frac{L_{\BB,0}^{\rm{new}}(z)\sin\delta_0(z)}{(z-s)z^n\vert \Omega_0(z) \vert}\right) .\label{equ:general solution}
\end{align}
For the phase shift $\delta_0(s)$, we take the parametrization in Ref.~\cite{Garcia-Martin:2011iqs}.
For the $\Omega_0(s)$ Omn\`es function, we take the $\Omega_{11}(s)$ matrix element of the coupled-channel Omn\`es matrix for the $\pi\pi$-$K\bar K$ $S$-wave interaction obtained in Ref.~\cite{Ropertz:2018stk}.  

The above equation provides a reasonable form that incorporates the $\pi\pi$ rescattering. The LHC part $L_{\BB,0}^{\rm{new}}(s)$ and the polynomial $P_{n-1}(s)$ may be determined by matching at low energies to the chiral amplitudes as done in Refs.~\cite{Donoghue:1996kw,Kang:2013jaa,Chen:2015jgl,Dong:2021lkh,Chen:2019hmz}. 
We perform the matching when the $\pi\pi$ rescattering is switched off, i.e., $\delta_0(s) = 0$, which leads to $\Omega_0(s)=1$. Consequently, for the $\BB\bar{\BB}\to\pi\pi$ process, we can approximate $L_{\BB,0}^{\rm{new}}(s)\sim\hat{A}_0^{\BB\ {\rm{new}}}(s)$ and $P_{n-1}(s) \sim A_0^{\BB\ {\rm{new}}}(s)$.  

Moreover, there is a polynomial ambiguity as discussed in Refs.~\cite{Anisovich:1996tx,Colangelo:2018jxw}. If the asymptotic value of the phase shift $\delta_0(s)$ is not 0 but $n\pi$ as $s\to \infty$, the corresponding Omn\` es function will approach $1/s^n$ asymptotically. In our case, the phase shift $\delta_0(s)\overset{s\to \infty}{\rightarrow} \pi$ implies $\Omega_0(s)\overset{s\to \infty}{\rightarrow} 1/s$, thus the general solution of the unitarity condition~\eqref{equ: RHC equation for solve} contains $3$ free parameters~\cite{Anisovich:1996tx,Colangelo:2018jxw} (assuming that $T^{\rm{new}}_{\BB\bar{\BB}\to\pi\pi,0}$ is asymptotically bounded by $s$). However, although the standard twice subtracted DR via Eq.~\eqref{equ:general solution} indeed grows like $s$ (notice that $n=2$), it contains only $2$ free parameters in the polynomial, i.e., one parameter less than the general solution. Hence we propose an oversubtracted DR (twice subtracted DR with an order-2 polynomial matching to the ChPT amplitudes)
that can be solved uniquely. 
In summary, the final DR is given as
\begin{align}
    T^{\rm{new}}_{\BB\bar{\BB}\to\pi\pi,0}(s)=\hat{A}_0^{\BB\ {\rm{new}}}(s)+\Omega_0(s)\left( A_0^{\BB\ {\rm{new}}}(s)+\frac{s^2}{\pi}\int_{4M_\pi^2}^{+\infty}{\rm{d}}z \frac{\hat{A}_0^{\BB\ {\rm{new}}}(z)\sin{\delta_0(z)}}{(z-s)z^2\vert \Omega_0(z) \vert} \right) .\label{equ:TBBbartopipi}
\end{align}

\begin{figure}[tb]
  \centering
      \begin{overpic}[width=1\linewidth]{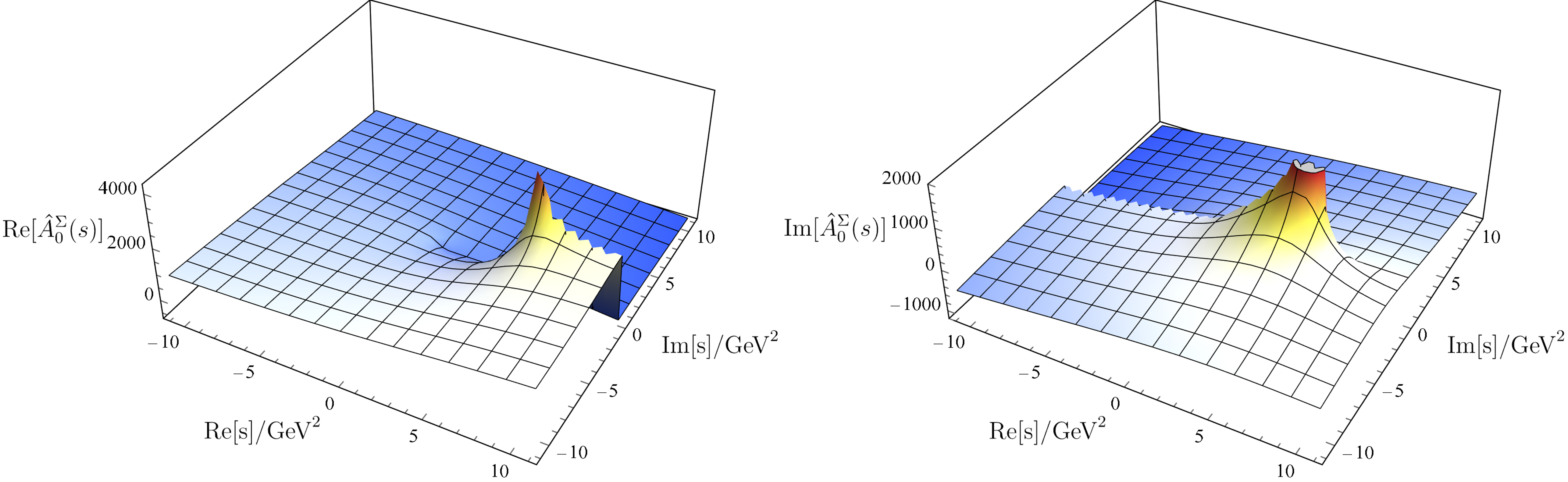}
      \end{overpic}
    \caption{Real (left panel) and imaginary (right panel) parts of the tree-level $t$- and $u$-channel exchange amplitude for the process of $\Sigma\bar{\Sigma}\to\pi\pi$ projected to the $\pi\pi$ $S$-wave as given in Eq.~(\ref{equ:LHC amplitude for Sigma}). The branch cut of the square root function is chosen to be along the positive real $s$ axis.} 
    \label{hatA0Sigma} 
  \end{figure}
  \begin{figure}[tb]
  \centering
      \begin{overpic}[width=1\linewidth]{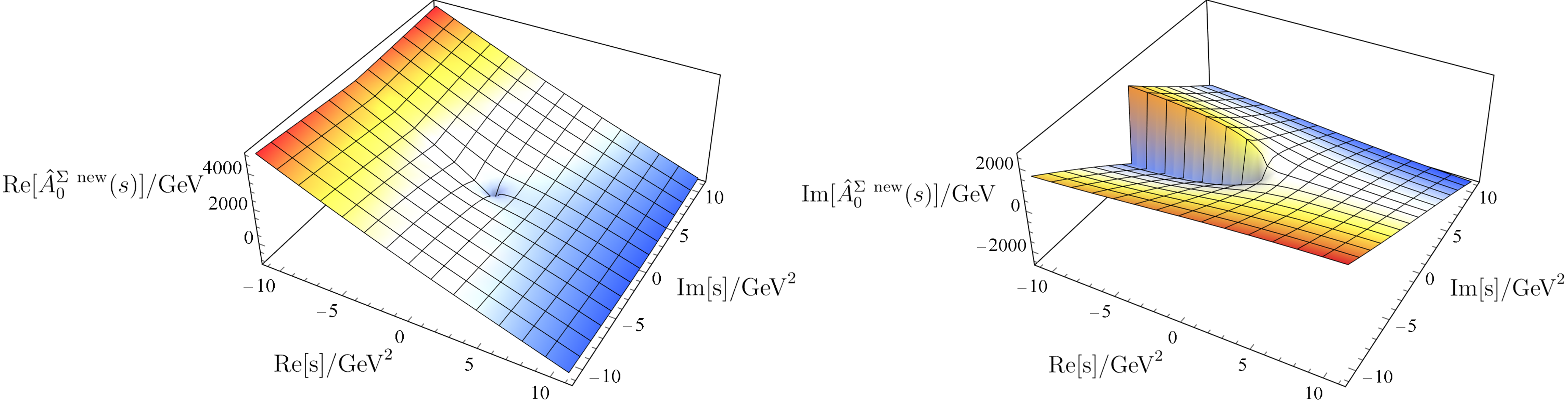}
      \end{overpic}
    \caption{Real (left panel) and imaginary (right panel) parts of the tree-level $t$- and $u$-channel exchange amplitude as given in Eq.~\eqref{equ:new defined LHC amplitude for Sigma} for the process of $\Sigma\bar{\Sigma}\to\pi\pi$ projected to the $\pi\pi$ $S$-wave. The amplitude is free of kinematical singularities and has only the desired LHC.
    } 
    \label{hatA0Sigmanew} 
  \end{figure}

From the above derivation, it is important to note that Eq.~(\ref{equ:TBBbartopipi}) can only be applied when the singularity of the LHC is exclusively included in $\hat{A}_0^{\BB\ {\rm{new}}}(s)$, and there is no overlap between the LHC and RHC. The original $t$- and $u$-channel exchange amplitudes Eqs.~(\ref{equ:LHC amplitude for N}-\ref{equ:LHC amplitude for Xi}) do not satisfy this condition due to the factor $\sqrt{s-4m_\BB^2}$. 
Let us take $\Sigma\bar \Sigma\to\pi\pi$ as an example.
From Fig.~\ref{hatA0Sigma}, it becomes apparent that the amplitude in Eq.~\eqref{equ:LHC amplitude for Sigma} includes the LHC $\left(-\infty,4M_\pi^2-{M_\pi^4}/{m_\Sigma^2}\right]$ derived from the particle exchanging in the crossed channel, as well as a kinematical cut in the physical region. 
Therefore, directly substituting Eq.~(\ref{equ:LHC amplitude for Sigma}) into Eq.~(\ref{equ:TBBbartopipi}) is invalid and disrupts the self-consistency of the theory. 
By employing the method described in Sec.~\ref{DR UR KS} to eliminate the kinematical singularities, the kinematical-singularity-free $S$-wave amplitude $\hat{A}^{\Sigma\ {\rm{new}}}_0(s)$ in Eq.~\eqref{equ:new defined LHC amplitude for Sigma} has only the LHC and satisfies the condition for Eq.~(\ref{equ:TBBbartopipi}), as demonstrated in Fig.~\ref{hatA0Sigmanew}.

At this stage, we can substitute the amplitude given by Eqs.~(\ref{equ:new defined LHC amplitude for N}-\ref{equ:new defined RHC amplitude for Xi}) into Eq.~(\ref{equ:TBBbartopipi}) to obtain the amplitude denoted as $T^{\rm{new}}_{\BB\bar{\BB}\to\pi\pi,0}(s)$. It includes the $S$-wave $\pi\pi$ rescattering and does not exhibit any kinematical singularities. Then, utilizing Eqs.~(\ref{equ:disc of Sigma}-\ref{equ:disc of N}), we get the discontinuity in Eq.~(\ref{DR integral}), and finally, the DR amplitude for $\BB\bar \BB\to \bar \BB \BB$ from exchanging correlated $S$-wave $\pi\pi$ is obtained by performing the dispersive integral.

\section{Determination of coupling constants}\label{numerical result}

Now we compare the two amplitudes, $\mathcal{M}^{\rm{OBE}}$ in Eq.~\eqref{equ:OBE amplitude} and $\mathcal{M}^{\rm{DR}}$ and Eq.~\eqref{DR integral}, to determine the coupling constant $g_{\BB\BB\sigma}$. 

\subsection{Matching $s$-channel amplitudes}\label{s-channel simulation}

\begin{figure}[tbhp]
\centering
\begin{overpic}[width=1\linewidth]{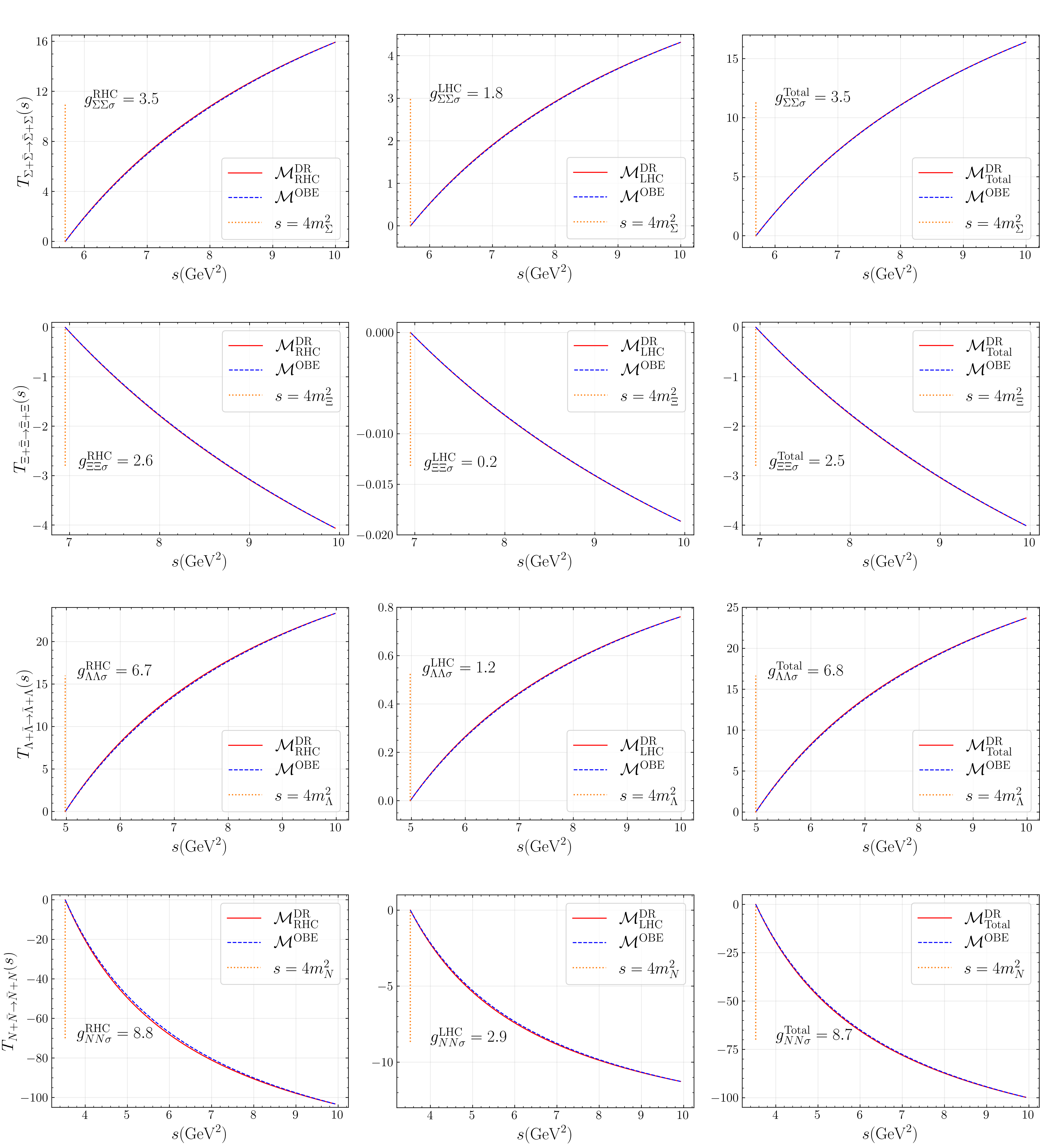}
    \end{overpic}
\caption{Comparison of the OBE amplitudes with different coupling constants obtained from $s$-channel $\sigma$ exchange and the DR amplitudes for different cases by using the central values of the LECs provided in Ref.~\cite{Ren:2012aj} and setting $\sqrt{s_0}$ to 0.8~GeV as in Ref.~\cite{Donoghue:2006rg}. The subscripts RHC, LHC and Total in $\mathcal{M}^{\rm{DR}}$ represent that the corresponding amplitudes consider only the RHC part shown in Fig.~\ref{FIG2} (c), only the LHC part  shown in Fig.~\ref{FIG2} (a) and (b), and both contributions combined, respectively.} 
  \label{FIG of amplitude} 
\end{figure}

Let us first compare the two amplitudes in Eqs.~(\ref{equ:OBE amplitude}, \ref{DR integral}) in the $s$-channel physical region, specifically $s\geq 4m_\BB^2$.
Since the amplitudes from exchanging $\sigma$ and from exchanging the correlated $S$-wave $\pi\pi$ have the same Lorentz structure, we can compare the two amplitudes at large $s$ values so that the pion masses and the $\sigma$ mass in the OBE amplitude play little role. A comparison of $\mathcal{M}^{\rm{OBE}}$ and $\mathcal{M}^{\rm{DR}}$ in the physical region of $s\geq 4m_\BB^2$ is shown in Fig.~\ref{FIG of amplitude}, where $g_{\BB\BB\sigma}$ has been adjusted so that the two amplitudes coincide in the physical region and $m_\sigma=0.5$~GeV is taken. 
In fact, matching Eqs.~(\ref{equ:OBE amplitude}, \ref{DR integral}) at $s\geq 4m_\BB^2$, one gets
\begin{align}
  -2\pi i C_\BB g_{\BB\BB\sigma}^2 \approx \int_{4M_\pi^2}^{s_0}\frac{{\rm{disc}}\left[ \mathcal{M}^{\rm{DR}}_{\BB\bar{\BB}\to\bar{\BB}\BB,0}(z)
\right]}{z-4m_\BB^2} \frac{s-m_\sigma^2}{s-z}{\rm{d}}z\,.\label{eq:MMcompare}
\end{align}
Since $s$ is much larger than both $m_
\sigma^2$ or $z\leq s_0 \simeq (0.8~\mathrm{GeV})^2$, one obtains the following sum rule:
\begin{align}
  g_{\BB\BB\sigma}^2 = - \frac1{2\pi i C_\BB} \int_{4M_\pi^2}^{s_0}\frac{{\rm{disc}}\left[ \mathcal{M}^{\rm{DR}}_{\BB\bar{\BB}\to\bar{\BB}\BB,0}(z)
\right]}{z-4m_\BB^2} {\rm{d}}z\,.\label{eq:sumrule}
\end{align}

The numerical results of the scalar coupling constants are presented in Table~\ref{couplings}, where the uncertainties in the second to fourth columns arise from the error propagated from those of the NLO LECs and the choice of the upper limit for the dispersive integral (see below), corresponding to Eqs.~(\ref{DR integral}, \ref{equ:TBBbartopipi}). In addition to the results obtained in the SU(3) framework, we also investigate  $g_{NN\sigma}$ in the SU(2) framework. The details are presented in Appendix~\ref{sec:AppSU2}, and the results are listed in the last row in Table~\ref{couplings}, labeled as $g^{\rm SU(2)}_{N N \sigma}$. 
Moreover, remarks are made on the difference in $g_{NN\sigma}$ under the SU(2) and SU(3) frameworks in Appendix~\ref{sec:AppSU2}.

Let us comment on the calculation of the two dispersive integrals. The first one, given by Eq.~(\ref{equ:TBBbartopipi}), is computed over the integration range of $[4M_\pi^2, (\sqrt{s_0}+\epsilon)^2]$. The second one, given by Eq.~(\ref{DR integral}), is integrated over $[(2M_\pi+\epsilon)^2, s_0]$.\footnote{Here, $\epsilon$ represents a small positive quantity that is relatively insignificant when compared to $\sqrt{s_0}$ and $2M_\pi$. As long as it is much smaller than $M_\pi$, the specific value has negligible impact on the results.} Note that the range of the second integral is completely covered by that of the first one to avoid unphysical singularities. 

The central values in Table~\ref{couplings} are obtained by setting $\sqrt{s_0}$ to 0.8~GeV as in Ref.~\cite{Donoghue:2006rg} and utilizing the central values of the NLO LECs provided in Ref.~\cite{Ren:2012aj}. The uncertainties of the NLO LECs as determined in Ref.~\cite{Ren:2012aj} are propagated to the coupling constants by using the bootstrap method. The resulting average values and corresponding standard deviations introduce the first source of errors in the third and fourth columns in Table~\ref{couplings} (the $b_i$ LECs appear only in the RHC contributions, and we have fixed the pion decay constant; thus the second column does not have errors from LECs). Furthermore, we vary $\sqrt{s_0}$ from 0.7~GeV to 0.9~GeV, which constitute the errors in the second column and the second source of errors in the third and fourth columns.  

\begin{table}[t]
\caption{The coupling constants $g_{\BB\BB\sigma}$ as given by the sum rule in Eq.~\eqref{eq:sumrule}.\footnote{The numerical results show that the total coupling $g_{\BB\BB\sigma}^{\rm{total}}$ does not align with the mere addition of the LHC and RHC couplings, $g_{\BB\BB\sigma}^{\rm{LHC}}+g_{\BB\BB\sigma}^{\rm{RHC}}$. This difference stems from the fact that both the LHC and RHC terms in Eq.~(\ref{equ:TBBbartopipi}) share the same phase factor, specifically $e^{i\delta_0(s)}$. Consequently, we anticipate the emergence of constructive and destructive interference effects in the subsequent computations involving the squared amplitude, as detailed in Eqs.~(\ref{equ:disc of Sigma}-\ref{equ:disc of N}), as well as during the integration procedures outlined in Eq.~(\ref{eq:sumrule}).} The second, third and fourth columns list the results  when only the LHC part shown in Fig.~\ref{FIG2} (a) and (b), only the RHC part shown in Fig.~\ref{FIG2} (c) and both of them are considered, respectively. The fifth to eleventh columns list the coupling constants from other references. For the seventh column, the values outside and inside the brackets represent the results calculated using different models in Ref.~\cite{Holzenkamp:1989tq}. The last column lists the mass (in MeV) of the $\sigma$ determined in the $t/u$-channel amplitude matching, as detailed in~\ref{t/u-channel simulation}. The last row for $g^{\rm SU(2)}_{N N \sigma}$ lists the results obtained in the SU(2) framework, as detailed in Appendix~\ref{sec:AppSU2}. 
}
 \centering
   \begin{tabular}{>{\centering\arraybackslash}p{3.0em} >{\centering\arraybackslash}p{3.5em} >{\centering\arraybackslash}p{6em} >{\centering\arraybackslash}p{6em}>{\centering\arraybackslash}p{2.5em}>{\centering\arraybackslash}p{2.5em}>
   {\centering\arraybackslash}p{4.5em}>{\centering\arraybackslash}p{2.5em}>{\centering\arraybackslash}p{2.5em}>{\centering\arraybackslash}p{2.5em}>{\centering\arraybackslash}p{2.5em}| >{\centering\arraybackslash}p{3.5em}}
     \toprule[2pt]
      & \text { LHC } & \text { RHC } & \text { Total } & \cite{Durso:1980vn} & \cite{Machleidt:1987hj} & \cite{Holzenkamp:1989tq}  & \cite{Reuber:1995vc}& \cite{Ronchen:2012eg} & \cite{Liu:2018bkx} & \cite{Zhao:2013ffn} & $m_\sigma$ \\
      \midrule[0.5pt]
      $g_{\Sigma \Sigma \sigma}$ & $1.8_{-0.5}^{+0.5}$ & $3.5_{-1.8-0.9}^{+2.0+0.8}$ & $3.5_{-1.3-0.4}^{+1.8+0.4}$ & - & -  & 10.85(8.92) & 4.65 & - &- &- & $519_{-48}^{+50}$\\
     $g_{\Xi \Xi \sigma}$ & $0.2_{-0.1}^{+0.1}$ & $2.6_{-1.4-0.6}^{+1.5+0.5}$ & $2.5_{-1.3-0.6}^{+1.5+0.5}$ & -& - & - & - & - & 3.4 & -& $614_{-81}^{+56}$\\
     $g_{\Lambda \Lambda \sigma}$ & $1.2_{-0.3}^{+0.4}$ & $6.7_{-1.1-1.7}^{+1.0+1.4}$ & $6.8_{-1.0-1.4}^{+1.0+1.1}$ & - &- & 8.18(6.54) & 4.37 & -  & -& 6.59 & $596_{-51}^{+41}$\\
     $g_{N N \sigma}$ & $2.9_{-0.8}^{+0.9}$ & $8.8_{-1.4-2.3}^{+1.4+1.9}$ & $8.7_{-1.3-1.4}^{+1.3+1.1}$&  \multirow{2}*{12.78} & \multirow{2}*{8.46} & \multirow{2}*{8.46} & \multirow{2}*{8.58} &  \multirow{2}*{13.85} &  \multirow{2}*{10.2} & \multirow{2}*{9.86} & $558_{-42}^{+33}$\\
     $g^{\rm SU(2)}_{N N \sigma}$ & $2.7_{-0.8}^{+0.8}$ & $12.5_{-0.2-3.2}^{+0.2+2.6}$ & $12.2_{-0.2-2.3}^{+0.2+1.9}$&   &  &  &  &   & & &$586_{-48}^{+38}$\\
    \bottomrule[2pt]
  \end{tabular}
  \label{couplings}
\end{table}

Results from other studies on these scalar couplings are also listed in Table~\ref{couplings}.
For $g_{NN\sigma}$ that has been estimated in many works, we find a good agreement with existing results, which supports the validity of our framework. Here we briefly discuss the methods used in the literature.
In Ref.~\cite{Durso:1980vn}, the authors investigated the $S$-wave $N\bar{N}\to\pi\pi$ amplitudes with the $\pi\pi$ rescattering and the results revealed that the intertwined contribution from the $\pi\pi$ $S$-wave can be elegantly described as a broad $\sigma$-meson with a mass of approximately $m_\sigma\!\sim\!4.8\,M_\pi$ and a coupling strength of $g_{NN\sigma}\!\sim\!12.78$. In Ref.~\cite{Machleidt:1987hj}, displaying the outcomes derived from the Bonn meson-exchange model, they found that the correlated $S$-wave $\pi\pi$ exchange can be further approximated by a zero width scalar exchange, with the corresponding mass and coupling constant readjusted to 550 MeV and 8.46, respectively. In Ref.~\cite{Holzenkamp:1989tq}, the authors also considered the $\sigma$ exchange as an effective parameterization for the correlated $S$-wave $\pi\pi$ exchange contribution. They utilized the result from the full Bonn meson-exchange model~\cite{Machleidt:1987hj} for the nucleon, i.e., the value in the sixth column of Table~\ref{couplings}, and $g_{\Lambda\Lambda\sigma}$ and $g_{\Sigma\Sigma\sigma}$ are determined by a fit to the empirical hyperon-nucleon data using two different models, with the distinction lying in whether higher-order processes involving a spin-$\frac{3}{2}$ baryon in the intermediate state were considered in the hyperon-nucleon interaction. 
In Ref.~\cite{Reuber:1995vc}, the authors calculated the $\BB\bar{\BB}'\to\pi\pi$ and $\BB\bar{\BB}'\to K\bar{K}$ amplitudes in the light of hadron-exchange picture. Based on an ansatz for Lagrangian, various symmetries and assumptions, they reduced the number of free parameters as many as possible, and then the parameters were fixed by adjusting the $N\bar{N}\to\pi\pi$ amplitudes to the quasi-empirical data. With these parameters and the existing $\pi\pi$ scattering phase shifts they got the $\BB\bar{\BB}'\to\pi\pi$ and $\BB\bar{\BB}'\to K\bar{K}$ amplitudes in the pseudo-physical region after solving the Blankenbecler-Sugar equation. Then employing the DR they got the spectral function which denotes the strength of a hadron-exchange process, namely the coupling constants. The eighth column in Table~\ref{couplings} represents their results, which are also similar to those reported in Ref.~\cite{Haidenbauer:2005zh}. 
In later development of the J\"ulich meson-exchange model in Ref.~\cite{Ronchen:2012eg}, the authors conducted an analysis of the coupled-channel dynamics and performed a simultaneous fit to the experimental data of various reactions, including $\pi N\to\pi N$, $\eta N$, $K\Lambda$ and $K\Sigma$, with the $\pi\pi N$ intermediate state parameterized as the $\sigma N$, $\pi\Delta$ and $\rho N$ channels. In their fitting, the coupling constant is determined to be $g_{NN\sigma}=13.85$. In Ref.~\cite{Zhao:2013ffn}, the authors used $g_{\Lambda\Lambda\sigma}=\frac{2}{3}g_{NN\sigma}$ from SU(3) consideration and took $g_{NN\sigma}$ from Ref.~\cite{Machleidt:1987hj}. 
In Ref.~\cite{Liu:2018bkx}, $g_{NN\sigma}={m_N}/{F_\pi}$ was determined using the linear $\sigma$ model~\cite{Gell-Mann:1960mvl}.
Then under the assumption that the $\sigma$ meson only couples to the $u$ and $d$ quarks, the authors got $g_{\Xi\Xi\sigma}=\frac{1}{3}g_{NN\sigma}$ based on the quark model consideration.
Additionally, in Ref.~\cite{Oset:2000gn}, the authors calculated the $NN$ potential arising from the exchange of a correlated $S$-wave isoscalar pion pair, i.e., the $\sigma$ channel, utilizing a unitary approach based on the lowest order chiral Lagrangian and the Bethe-Salpeter equation for the analysis of $\pi\pi$ scattering. 
A qualitative estimate for $g_{NN\sigma}\sim 5$ was obtained, at the right order of the values quoted in Table~\ref{couplings}.

\subsection{Matching \texorpdfstring{$\bm{t/u}$}{t/u}-channel amplitudes}\label{t/u-channel simulation}

\begin{figure}[tb]
\centering
    \begin{overpic}[width=0.8\linewidth]{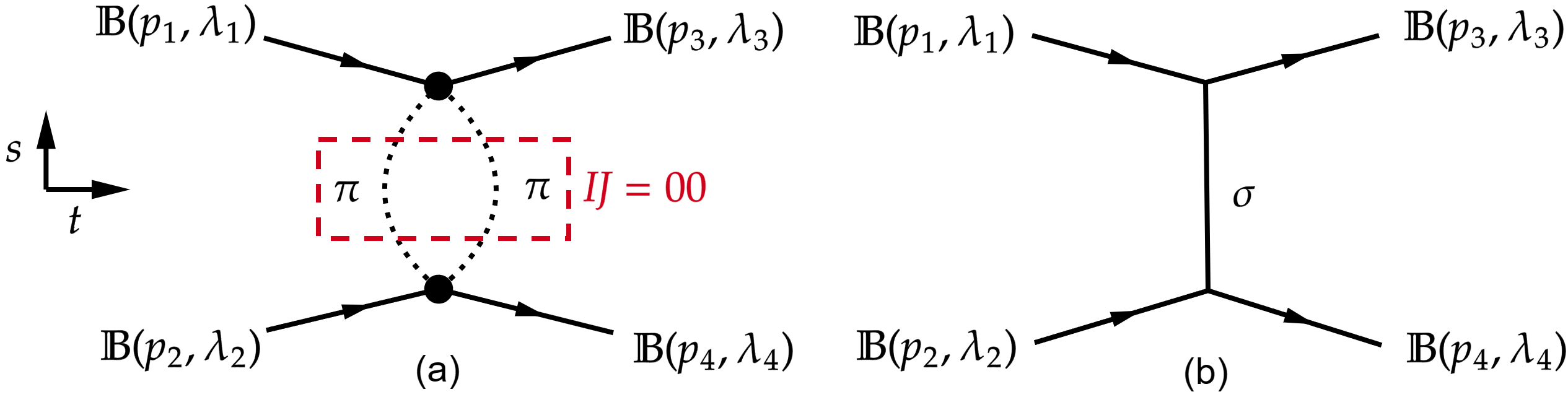}
    \end{overpic}
\caption{The Feynman diagram for the $t$-channel process of $\BB\BB\to \BB\BB$ with the intermediate state of $\pi\pi$ (a) or $\sigma$ (b). In (a), the black dots imply $\pi\pi$ interaction.}
  \label{t-channel Feynman diagram} 
\end{figure}

In the preceding subsection, it becomes evident that for the $s$-channel process of $\BB\bar{\BB}\to\bar{\BB}\BB$, the selection of an apt coupling constant $g_{\BB \BB \sigma}$ allows for the $\sigma$ exchange to mimic the correlated $\pi\pi$ intermediate state with $IJ=00$ in physical region, $s\geq 4m_\BB^2$. However, when employing the OBE model to estimate the interaction between hadrons, the $\sigma$ is exchanged in the $t$- or $u$-channel, as illustrated in Fig.~\ref{t-channel Feynman diagram}~(b) rather than in the $s$-channel, as demonstrated in Fig.~\ref{FIG1}~(b). Therefore, to derive the parameters for the $\sigma$ exchange that can be used in the OBE model, one needs to conduct an analysis of the $t$- and $u$-channel meson-exchange processes. 
As elaborated in Appendix~\ref{crossing relation appendix}, the crossing symmetry relations provide a means to relate the $t(u)$-channel process to the $s$-channel one. It becomes evident that, should we manage to align the two amplitudes within the non-physical region of the $s$-channel process, specifically $s\in[4m_\BB^2-t,0]$, we can subsequently match the corresponding pair of amplitudes within the physical region of the $t(u)$-channel process, i.e., $t\geq4m_\BB^2$, relevant for the low-energy $\BB\BB$ scattering.  

\begin{figure}[h]
\centering
    \begin{overpic}[width=1\linewidth]{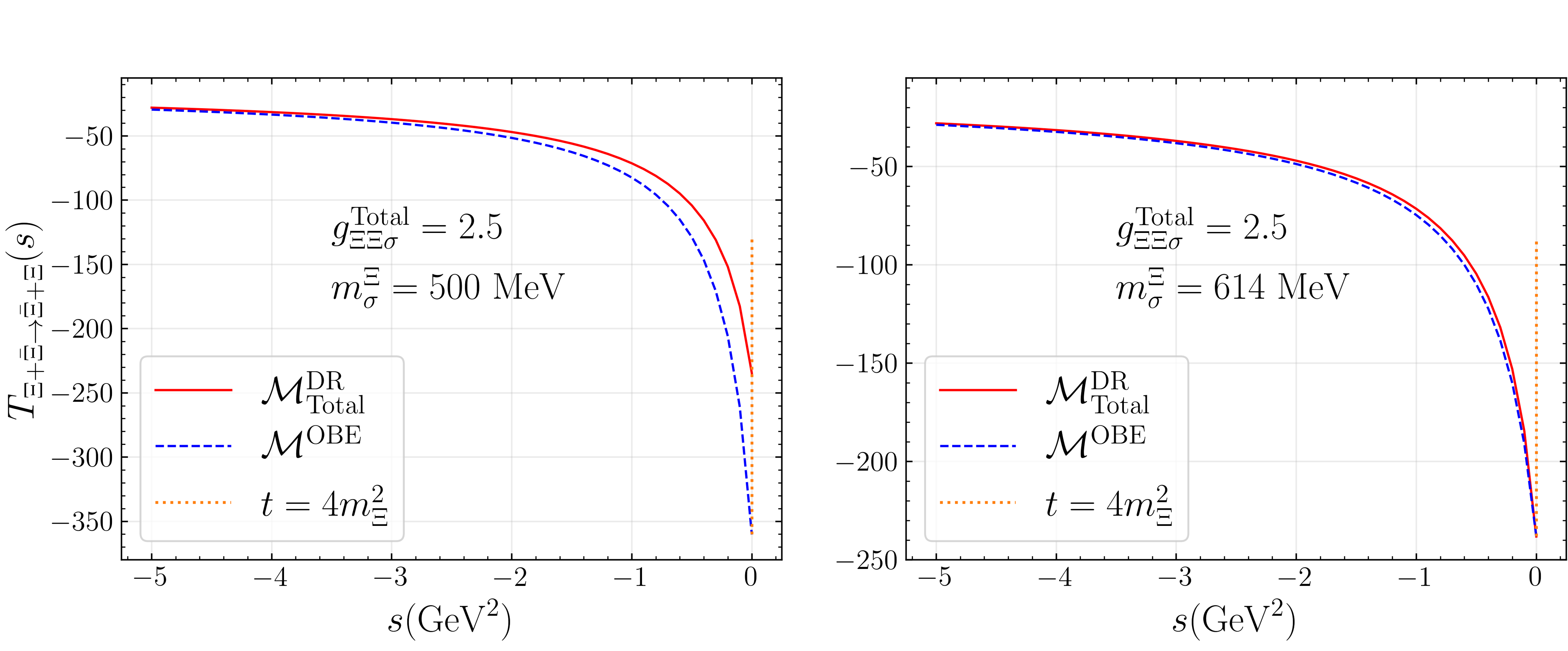}
    \put(42,32){\normalsize{(a)}}
    \put(92,32){\normalsize{(b)}}
    \end{overpic}
  \caption{Comparison of the OBE amplitude, with the coupling constant taking the central value listed in Table~\ref{couplings} and different $m_\sigma^\Xi$ values in the process of $\Xi \bar{\Xi}\to\bar{\Xi}\Xi$, and the DR amplitude using the central values of the LECs provided in Ref.~\cite{Ren:2012aj} and setting $\sqrt{s_0}$ to 0.8~GeV as in Ref.~\cite{Donoghue:2006rg}.} 
  \label{t-channel compare1}
\end{figure}

In order for the $\sigma$ exchange to approximate the $S$-wave correlated two pions in the few hundred MeV region, we also need to adjust the $\sigma$ mass in addition to the couplings derived above.\footnote{Since in the $\BB \BB$ scattering physical region, the exchanged two pions cannot go on shell, a real mass, instead of the complex pole, for the $\sigma$ meson in the OBE model should be used. }
As an example, in Fig.~\ref{t-channel compare1}, we show the comparison of the OBE amplitude and the DR amplitude for the $\Xi\Xi$ case at the $t$-channel threshold. One finds from Fig.~\ref{t-channel compare1}~(b) that by adjusting the $\sigma$ mass to about $614_{-81}^{+56}$~MeV, the DR amplitude using the central values of the LECs can be very well reproduced. The matching point has been chosen to be $s=0$~GeV$^2$, corresponding to the $t$-channel $\BB\BB$ threshold.
To see the dependence on the $\sigma$ mass, we also show the comparison for $m_\sigma=500$~MeV in Fig.~\ref{t-channel compare1}~(a).

\begin{figure}[h]
\centering
\begin{overpic}[width=1\linewidth]{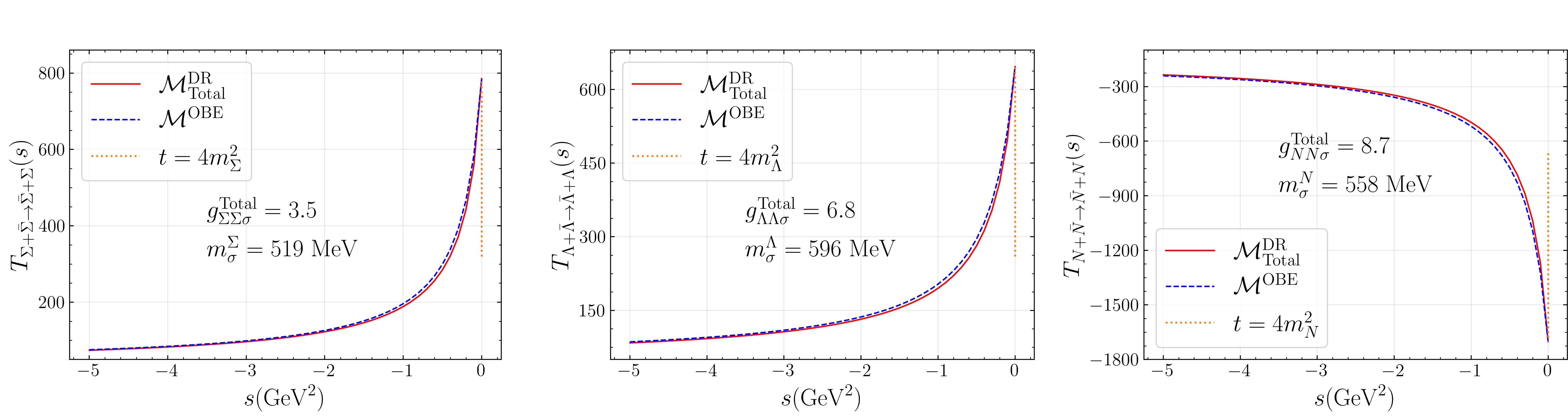}
    \end{overpic}
\caption{Matching at $\BB\BB$ threshold the OBE amplitudes, with the coupling constant taking the central value listed in Table~\ref{couplings}, to the DR amplitudes using the central values of the LECs provided in Ref.~\cite{Ren:2012aj} and setting $\sqrt{s_0}$ to 0.8~GeV as in Ref.~\cite{Donoghue:2006rg}.} 
  \label{t-channel total picture} 
\end{figure}

The aforementioned analysis can be readily extended to the other ground state octet baryons, yielding the results shown in Fig.~\ref{t-channel total picture}. From Figs.~\ref{FIG of amplitude},~\ref{t-channel compare1} (b) and~\ref{t-channel total picture}, it is apparent that if our aim is to use a simple $\sigma$ exchange in the OBE model to concurrently match a complex correlated $\pi\pi$ exchange with $IJ=00$ in the $s$-, $t$- and $u$-channel physical region, the $m_\sigma$ values required by different processes differ. Specifically, we find $m_\sigma^{\Sigma}=519_{-48}^{+50}$ MeV, $m_\sigma^{\Xi}= 614_{-81}^{+56}$ MeV, $m_\sigma^{\Lambda}= 596_{-51}^{+41}$ MeV and $m_\sigma^{N}= 558_{-42}^{+33}$ MeV, where the uncertainties correspond to those of the couplings added in quadrature.\footnote{The superscript of $m_\sigma^\BB$ is utilized to represent the mass of this $\sigma$ which is derived from the process of $\BB\bar{\BB}\to\bar{\BB}\BB$.} These values are listed in the last column of Table~\ref{couplings}. This echoes previous attempts to modify the mass of $\sigma$, a broad resonance with a mass approximately equal to $4.8\ M_\pi$~\cite{Durso:1980vn}, to a mass of 550 MeV with a zero width~\cite{Machleidt:1987hj}, which is within all the above ranges. The goal of such modification was to allow a single $\sigma$ exchange to more accurately replicate the results of a correlated $\pi\pi$ exchange with $IJ=00$. 

\section{summary}\label{summary}

In this work, we evaluate the couplings of the $\sigma$ meson to the  $\frac{1}{2}^+$ ground state light baryons, which are essential inputs of the OBE models, by matching the baryon-baryon scattering amplitudes through correlated $S$-wave isoscalar $\pi\pi$ intermediate state to the OBE ones. 
Using the LO and NLO SU(3) chiral baryon-meson Lagrangians, we carefully handle the kinematical singularities and utilize DR and incorporate the $\pi\pi$ rescattering by Muskhelishvili-Omn\`es representation to obtain the DR amplitude. 
Considering the phenomenological $\sigma$ exchange as an effective parameterization for the correlated $\pi\pi$ exchange contribution in the $IJ=00$ channel, we determine the scalar coupling constants $g_{\BB\BB\sigma}$ from the $s$-channel matching, as listed in Table~\ref{couplings}. Specifically, $g_{\Sigma\Sigma\sigma}=3.5_{-1.3}^{+1.8}$, $g_{\Xi\Xi\sigma}=2.5_{-1.4}^{+1.5}$, $g_{\Lambda\Lambda\sigma}=6.8_{-1.7}^{+1.5}$, and $g_{NN\sigma}=8.7_{-1.9}^{+1.7}$, where the errors are obtained by adding the corresponding ones in Table~\ref{couplings} in quadrature.
This is achieved by comparing the DR amplitude and OBE amplitude in the physical region of the $s$-channel process, specifically, $s\geq 4m_\BB^2$. Concurrently, we estimate the uncertainties of the scalar coupling constants arising from the NLO LECs~\cite{Ren:2012aj} and variation of the upper limit for the dispersive integral.  
Moreover, by extending the analysis to the physical region of the corresponding $t/u$-channel process via the crossing relation, we obtain the $\sigma$ mass to be used together with the determined $\BB\BB\sigma$ coupling constant. The value depends on the process  but is always around 550~MeV. 
We also compute the $NN\sigma$ coupling by matching to the SU(2) CHPT amplitude with the LECs determined in Refs.~\cite{Hoferichter:2015hva,Hoferichter:2015tha}, and the result is $g_{NN\sigma}^{\rm SU(2)}=12.2_{-2.3}^{+1.9}$.

The effective coupling constants obtained here can be used to describe the interaction between light hadrons and other hadrons through the $\sigma$ exchange. The same method can be applied to the determination of the coupling constants of $\sigma$ and other hadrons, such as heavy mesons and baryons, the interactions between which are crucial to understand the abundance of exotic hadron candidates observed at various experiments in last two decades.

\begin{acknowledgements}
We would like to thank Ulf-G. Mei{\ss}ner for a careful reading of the manuscript.
This work is supported in part by the Chinese Academy of Sciences under Grants No.~XDB34030000 and No. YSBR-101; by the National Natural Science Foundation of China (NSFC) and the Deutsche Forschungsgemeinschaft (DFG) through the funds provided to the Sino-German Collaborative Research Center TRR110 ``Symmetries and the Emergence of Structure in QCD'' (NSFC Grant No. 12070131001, DFG Project-ID 196253076); by the NSFC under Grants No. 12125507, No. 11835015, and No. 12047503; and by the Postdoctoral Fellowship Program of China
Postdoctoral Science Foundation (CPSF) under No. GZC20232773 and the CPSF No. 2023M743601.
\end{acknowledgements}

\begin{appendix}
    
\section{Isospin conventions}\label{SecI}
In this work, we use the following isospin conventions~\cite{deSwart:1963pdg}:
\begin{align}
\vert\pi^+\rangle&=-\vert 1,1\rangle , & \vert\pi^0\rangle&=\vert 1,0\rangle ,\notag\\
\vert\pi^-\rangle&=\vert 1,-1\rangle , &
\vert \Sigma^+\rangle&=-\vert 1,1\rangle ,\notag\\ 
\vert\Sigma^0\rangle&=\vert 1,0\rangle , & \vert\Sigma^-\rangle&=\vert 1,-1\rangle ,\notag\\
\vert\bar{\Sigma}^+\rangle&=-\vert 1,1\rangle , & \vert\bar{\Sigma}^0\rangle&=\vert 1,0\rangle ,\notag\\ 
\vert\bar{\Sigma}^-\rangle&=\vert 1,-1\rangle , &
\vert\Xi^0\rangle&=\vert\frac{1}{2},\frac{1}{2}\rangle ,\notag\\ 
\vert\Xi^-\rangle&=\vert\frac{1}{2},-\frac{1}{2}\rangle , &
\vert\bar{\Xi}^+\rangle&=-\vert\frac{1}{2},\frac{1}{2}\rangle ,\notag\\ \vert\bar{\Xi}^0\rangle&=\vert\frac{1}{2},-\frac{1}{2}\rangle , & \vert\Lambda^0\rangle&=\vert 0,0\rangle ,\notag\\
\vert p\rangle&=\vert\frac{1}{2},\frac{1}{2}\rangle , & \vert n\rangle&=\vert\frac{1}{2},-\frac{1}{2}\rangle ,\notag\\
\vert\bar{n}\rangle&=\vert\frac{1}{2},\frac{1}{2}\rangle , & |\bar{p}\rangle&=-\vert\frac{1}{2},-\frac{1}{2}\rangle .\notag
\end{align}
Therefore, we can readily obtain the isoscalar state $\vert I=0, I_3=0 \rangle$ composed of $\pi\pi$, $\BB\bar{\BB}$ and $\bar{\BB}\BB$ in the particle basis. 

\section{$g_{NN\sigma}$ from SU(2) ChPT}\label{sec:AppSU2}
It is worth mentioning that in the context of $\pi N$ interaction,  it is more common to utilize the Lagrangian within the SU(2) framework, the LO Lagrangian is given by
\begin{align}
\mathcal{L}_{\pi N}^{(1)}=&\bar{\Psi} \left( i \mathcal{D} \mkern -9.5 mu /-m_N + \frac{g_A}{2}\gamma^\mu\gamma_5u_\mu \right) \Psi ,\label{equ: SU(2) LO Lagrangian}
\end{align}
where $g_A$ represents the nucleon axial-vector coupling constant in the chiral limit and is related to the SU(3) LECs via $g_A=D+F$. At the NLO, 
\begin{align}
\mathcal{L}_{\pi N}^{(2)}= & c_1 \operatorname{Tr}\left(\chi_{+}\right) \bar{\Psi} \Psi-\frac{c_2}{4 m_N^2} \operatorname{Tr}\left(u_\mu u_\nu\right)\left(\bar{\Psi} \mathcal{D}^\mu \mathcal{D}^\nu \Psi+\text {H.c.}\right)\notag \\
& +\frac{c_3}{2} \operatorname{Tr}\left(u^\mu u_\mu\right) \bar{\Psi} \Psi-\frac{c_4}{4} \bar{\Psi} \gamma^\mu \gamma^\nu\left[u_\mu, u_\nu\right] \Psi+c_5 \bar{\Psi}\left[\chi_{+}-\frac{1}{2} \operatorname{Tr}\left(\chi_{+}\right)\right] \Psi \notag \\
& +\bar{\Psi} \sigma^{\mu \nu}\left[\frac{c_6}{2} f_{\mu \nu}^{+}+\frac{c_7}{2} v_{\mu \nu}^{(s)}\right] \Psi ,\label{equ: SU(2) NLO Lagrangian}
\end{align}
which contains seven LECs $c_i$~\cite{Fettes:1998ud,Fettes:2000gb,Scherer:2012xha,Meissner:2022cbi}, the first four of which are determined in Refs.~\cite{Hoferichter:2015tha,Hoferichter:2015hva} as (in units of ${\rm{GeV}}^{-1}$),
\begin{align}
c_1=-0.74\pm0.02 , \quad c_2=1.81\pm0.03 , \quad c_3=-3.61\pm0.05 , \quad c_4=2.17\pm0.03 .\label{LECs for SU(2)}
\end{align}
By utilizing the Eqs.~(\ref{equ: SU(2) LO Lagrangian}, \ref{equ: SU(2) NLO Lagrangian}) and the above LECs, we obtain the following results through the $s$-channel matching as detailed in Sect.~\ref{s-channel simulation}:
\begin{align}
g_{NN\sigma}^{\rm{LHC}} = {2.7_{-0.8}^{+0.8}} , \quad
g_{NN\sigma}^{\rm{RHC}} =
{12.5 _{-0.2-3.2}^{+0.2+2.6}} ,\quad
g_{NN\sigma}^{\rm{Total}} =
{12.2 _{-0.2-2.3}^{+0.2+1.9}} .\label{SU(2) c3=-4.7}
\end{align}
Notice that here for consistency with the $c_i$ values, we take $F_\pi=92.2$~MeV {and $g_A=1.2723$} used in Refs.~\cite{Hoferichter:2015tha,Hoferichter:2015hva}, larger than the value used in the main text.
Meanwhile, from matching the $t/u$-channel amplitudes, we find $m_\sigma^{N\ {\rm{SU(2)}}}= {586_{-48}^{+38}}$~MeV.
The $g_{NN\sigma}^\text{Total}$ value geiven above is close to the real part of the coupling defined as the residue of the $\pi\pi \to N \bar N$ amplitude at the $f_0(500)$ pole obtained in Ref.~\cite{Hoferichter:2023mgy}, which is $12.1\pm1.4$.

\begin{figure}[h]
\centering
      \includegraphics[width=0.618\linewidth]{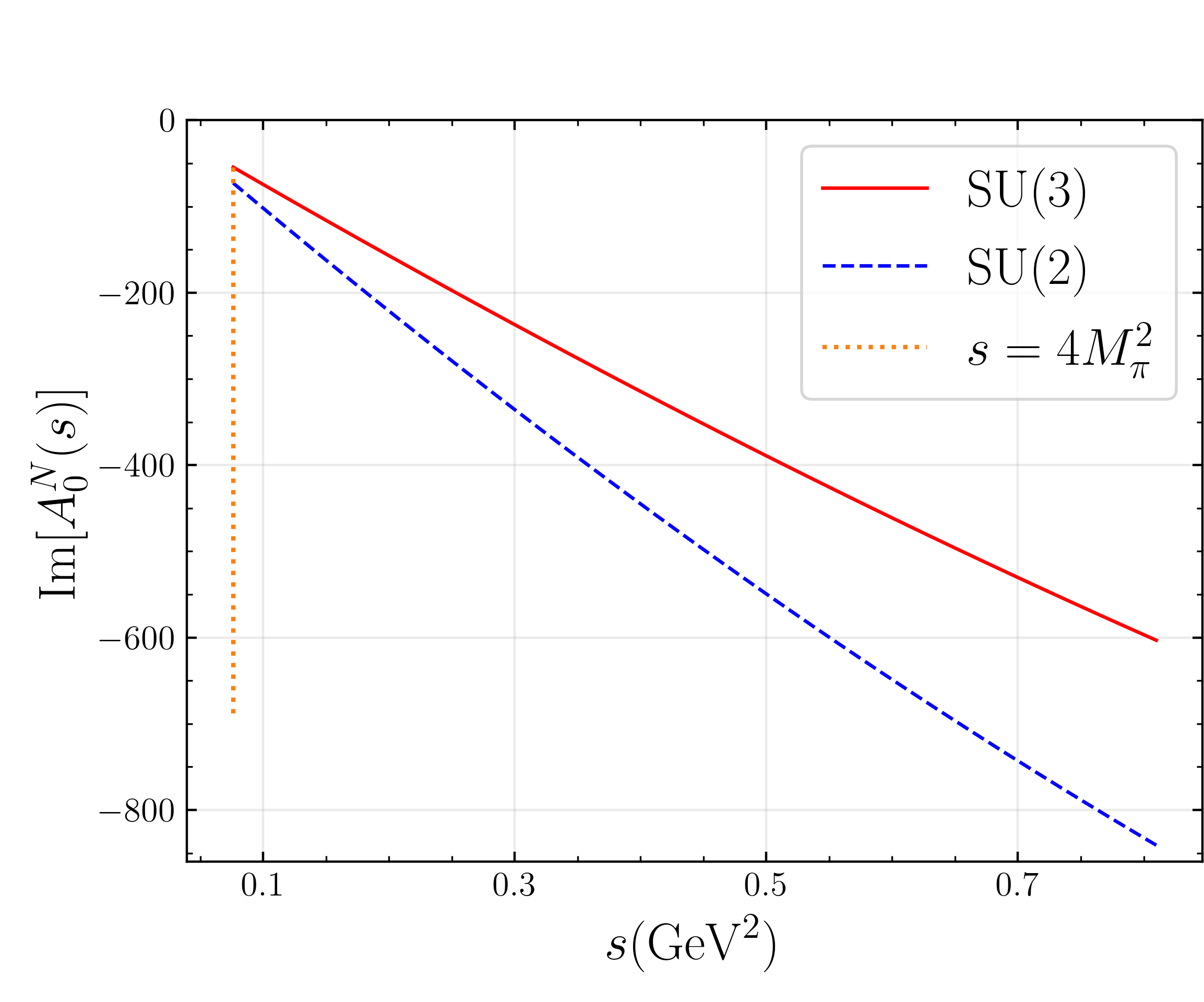}
  \caption{The contact term amplitudes of the $N\bar{N}\to\pi\pi$ process derived from the SU(2) and SU(3) chiral Lagrangians using the central values of LECs determined in Refs.~\cite{Hoferichter:2015tha,Hoferichter:2015hva} and Ref.~\cite{Ren:2012aj}, respectively.
  } 
  \label{NLO contact term amplitude for SU(2) SU(3)}
\end{figure}

As per Table~\ref{couplings}, the $g_{NN\sigma}^{\rm{RHC}}$ central value calculated using the ChPT NLO Lagrangian within the SU(2) framework deviates from its value within the SU(3) framework. 
In Fig.~\ref{NLO contact term amplitude for SU(2) SU(3)}, we show a comparison of the $S$-wave tree-level amplitudes of the contact terms for the $N\bar{N}\to\pi\pi$ process from the SU(3) chiral Lagrangian with that from the SU(2) chiral Lagrangian, the LECs of which are taken from Ref.~\cite{Ren:2012aj} and Refs.~\cite{Hoferichter:2015tha,Hoferichter:2015hva}, respectively.
One sees a clear deviation. We have checked that the deviation from the SU(2) result would be larger if we use the central values of the SU(3) LECs determined by other groups~\cite{Borasoy:1996bx,Ikeda:2012au,Guo:2012vv,Mai:2012dt}.
Nevertheless, the values of $g_{NN\sigma}$ and $g_{NN\sigma}^\text{SU(2)}$ from RHC contributions agree within uncertainties. One notices that Refs.~\cite{Ren:2012aj} and Refs.~\cite{Hoferichter:2015tha,Hoferichter:2015hva} considered different experimental and lattice data sets.

\section{The crossing relation}\label{crossing relation appendix}

Based on the crossing symmetry, we can establish a relation between the $s$-channel helicity amplitude of $\BB\bar{\BB}\to\bar{\BB}\BB$ and the $t$-channel helicity amplitude of $\BB\BB\to \BB\BB$ or the $u$-channel helicity amplitude of $\BB\bar{\BB}\to\bar{\BB}\BB$. Using crossing symmetry relations for systems with spin~\cite{Hara:1970gc,Martin:1970hmp,Hebbar:2020ukp},\footnote{In the context of crossing relation, for a $t$-channel process of $\BB(p_1)+\BB(p_2)\to \BB(p_3)+\BB(p_4)$, as illustrated in Fig.~\ref{t-channel Feynman diagram}, $s$ refers to $(p_1-p_3)^2$ while $t$ refers to $(p_1+p_2)^2$.}  the amplitude for the $t$-channel process of $\BB\BB\to \BB\BB$ via the correlated $\pi\pi$ intermediate state with $IJ=00$ can be expressed as
\begin{align}
    \mathcal{M}_{\BB(\lambda_1)\BB(\lambda_3)\to \BB(\lambda_2)\BB(\lambda_4),0}^{t\text{-channel}}&(t,s)\notag\\
    =\sum\limits_{\lambda_i'}&d^{\frac{1}{2}}_{\lambda_1\lambda_1'}(\alpha_1)d^{\frac{1}{2}}_{\lambda_2\lambda_2'}(\alpha_2)d^{\frac{1}{2}}_{\lambda_3\lambda_3'}(\alpha_3)d^{\frac{1}{2}}_{\lambda_4\lambda_4'}(\alpha_4)\mathcal{M}_{\BB(\lambda_1')\bar{\BB}(\lambda_2')\to\bar{\BB}(\lambda_3')\BB(\lambda_4'),0}^{s\text{-channel}}(s) ,\label{crossing relation}
\end{align}
where $\alpha_i$ represents the Wigner rotation angles  corresponding to the Lorentz transformation from the $s$-channel c.m. frame to the $t$-channel c.m. frame, and the subscript $0$ signifies that the $\pi\pi$ of either the $t$-channel process or the $s$-channel process forms an isoscalar $S$-wave. Considering that the crossing relation, Eq.~(\ref{crossing relation}), is solely dependent on the particles of the external lines, the same relation is applicable regardless of whether there is a $\sigma$ exchange or a correlated $\pi\pi$ exchange, namely,
\begin{align}
    \raisebox{-0.5\height}{\includegraphics[width=3cm]{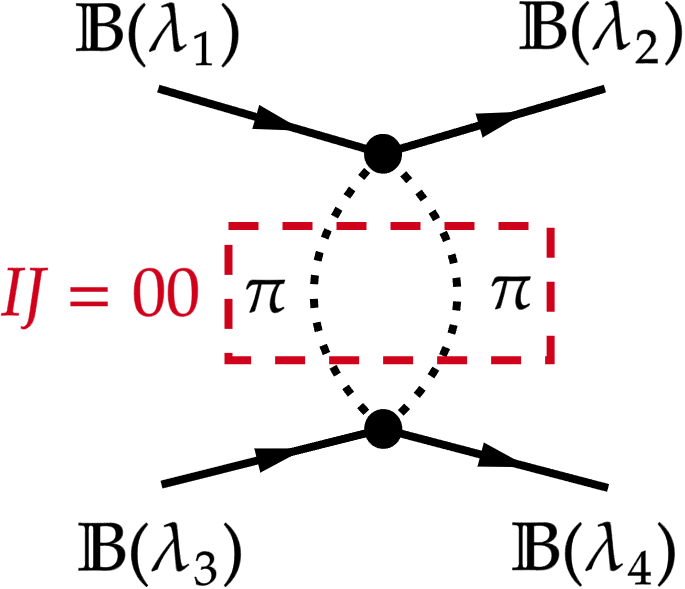}}&=\sum\limits_{\lambda_i'}d^{\frac{1}{2}}_{\lambda_1\lambda_1'}(\alpha_1)d^{\frac{1}{2}}_{\lambda_2\lambda_2'}(\alpha_2)d^{\frac{1}{2}}_{\lambda_3\lambda_3'}(\alpha_3)d^{\frac{1}{2}}_{\lambda_4\lambda_4'}(\alpha_4)\raisebox{-0.5\height}{\includegraphics[width=3cm]{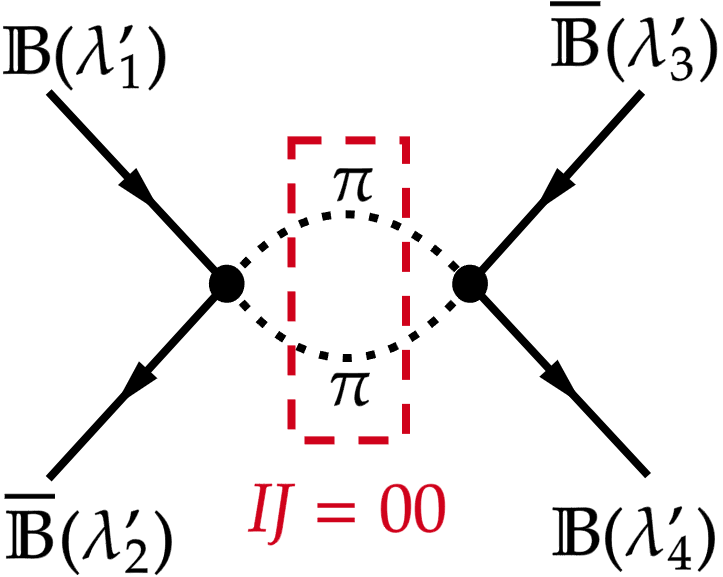}} ,\label{equ:t =Sum s for pipi}\\ \raisebox{-0.5\height}{\includegraphics[width=2.7cm]{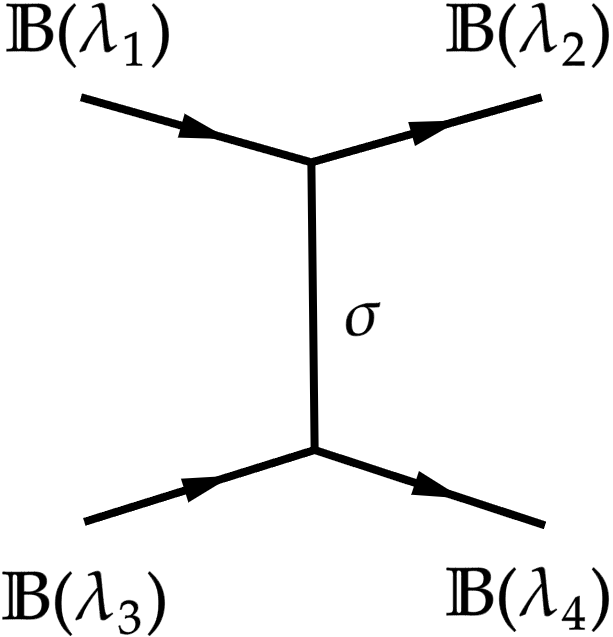}}&=\sum\limits_{\lambda_i'}d^{\frac{1}{2}}_{\lambda_1\lambda_1'}(\alpha_1)d^{\frac{1}{2}}_{\lambda_2\lambda_2'}(\alpha_2)d^{\frac{1}{2}}_{\lambda_3\lambda_3'}(\alpha_3)d^{\frac{1}{2}}_{\lambda_4\lambda_4'}(\alpha_4)\raisebox{-0.5\height}{\includegraphics[width=3cm]{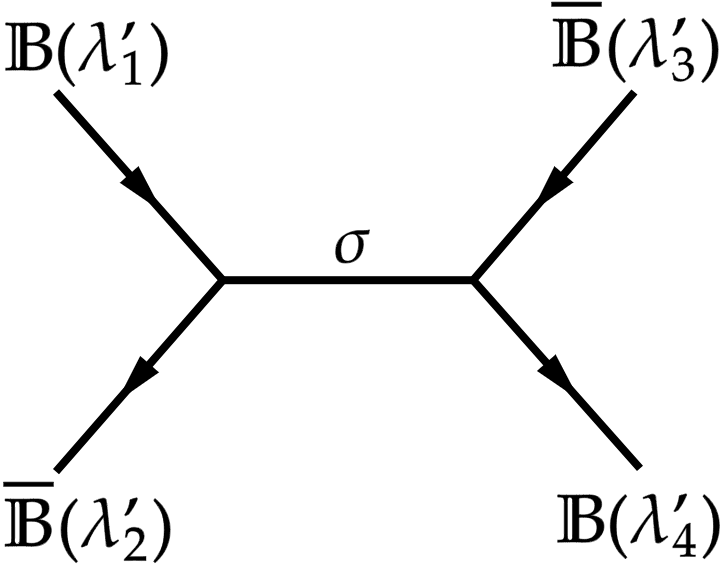}} .\label{equ:t =Sum s for sigma}
\end{align}
We then obtain 
\begin{align}
    &\raisebox{-0.5\height}{\includegraphics[width=3cm]{fy1.png}}-\raisebox{-0.5\height}{\includegraphics[width=2.7cm]{fy3.png}}\notag\\
    &=\sum\limits_{\lambda_i'}d^{\frac{1}{2}}_{\lambda_1\lambda_1'}(\alpha_1)d^{\frac{1}{2}}_{\lambda_2\lambda_2'}(\alpha_2)d^{\frac{1}{2}}_{\lambda_3\lambda_3'}(\alpha_3)d^{\frac{1}{2}}_{\lambda_4\lambda_4'}(\alpha_4)\left(\raisebox{-0.5\height}{\includegraphics[width=3cm]{fy2.png}}-\raisebox{-0.5\height}{\includegraphics[width=3cm]{fy4.png}}\right) .\label{equ:50 for t-channel simulation}
\end{align}

Since our goal is to ensure that the amplitude of the correlated $\pi\pi$ exchange with $IJ=00$ and that of the $\sigma$ exchange are approximately the same for the $t$-channel process of $\BB\BB\to \BB\BB$ within the $t$-channel physical region, i.e., $t\geq 4m_\BB^2$, 
we require the corresponding $s$-channel amplitudes of $\BB\bar{\BB}\to \bar{\BB}\BB$ to approximate each other as well as possible, i.e.,
\begin{align}
\mathcal{M}_{\BB\bar{\BB}\to\bar{\BB}\BB}^{\rm{OBE}}(s)\simeq \mathcal{M}_{\BB\bar{\BB}\to\bar{\BB}\BB,0}^{\rm{DR}}(s)\label{t-channel simulation condition}
\end{align}
when $s\in[4m_\BB^2-t,0]$. The $u$-channel process mirrors this exactly.

\end{appendix}

\bibliography{ref}

\end{document}